\documentclass[twocolumn]{aastex7}
\shortauthors{Byun et al.}
\received{June 22, 2026}
\revised{October 4, 2026}
\accepted{October 25, 2026}
\submitjournal{ApJ}
\graphicspath{{./}{figures/}}

\begin{document}

\title{Linking Detectable Tidal Features to Stellar Halo Growth: An Integrated Analysis of the Outskirts of Massive Early-type Galaxies}

\author[orcid=0000-0002-7762-7712, gname=Woowon, sname=Byun]{Woowon Byun}
\affiliation{Korea Astronomy and Space Science Institute, Daejeon 34055, Republic of Korea}
\email[show]{wbyun87@gmail.com}

\author[orcid=0000-0003-0134-8968, gname=Yongmin, sname=Yoon]{Yongmin Yoon} 
\affiliation{Department of Astronomy and Atmospheric Sciences, Kyungpook National University, Daegu 41566, Republic of Korea}
\email[show]{yyoon@knu.ac.kr}

\author[orcid=0000-0001-9544-7021, gname=Kyungwon, sname=Chun]{Kyungwon Chun} 
\affiliation{Korea Astronomy and Space Science Institute, Daejeon 34055, Republic of Korea}
\email{kwchun@kasi.re.kr}

\author[orcid=0000-0002-9434-5936, gname=Jongwan, sname=Ko]{Jongwan Ko} 
\affiliation{Korea Astronomy and Space Science Institute, Daejeon 34055, Republic of Korea}
\affiliation{Department of Astronomy and Space Science, University of Science and Technology, Daejeon 34113, Republic of Korea}
\email{jwko@kasi.re.kr}

\author[orcid=0000-0002-1710-4442, gname=Jae-Woo, sname=Kim]{Jae-Woo Kim}
\affiliation{Korea Astronomy and Space Science Institute, Daejeon 34055, Republic of Korea}
\email{kjw0704@kasi.re.kr}

\begin{abstract}

Motivated by the finding that massive galaxies with tidal features (GwT) are preferentially found in low-density environments, one wonders whether this trend arises from enhanced debris survivability or from more active mass-assembly histories. If driven by the latter, structural signatures, such as more developed stellar halos, should also form through cumulative accretion. To address this, we analyze azimuthally averaged surface-brightness profiles of 436 massive early-type galaxies from the DESI Legacy Imaging Survey, comparing the median-stacked profiles of galaxies with tidal features to those without. We find that, after resampling by stellar mass and environmental density, a subtle but systematic structural divergence emerges in the galaxy outskirts ($r>5\times r_{50}$). Specifically, the GwT population in the lower-density \& higher-mass regime exhibits a relative intensity excess. This suggests that more developed stellar halos have formed, likely due to longer and/or more significant accretion events. In contrast, those in the higher-density \& lower-mass regime show a relative deficiency in intensity, which is reasonable to interpret as a result of structural disruption from environmental effects rather than a lack of cumulative accretion. Our findings suggest that visible tidal features---particularly in massive systems within quiet environments---do not merely reflect recent stochastic accretion events but also hint at the long-term assembly history of galaxies. This work is a depth-limited study that offers a glimpse into an integrated analysis of tidal debris and stellar halos, thereby introducing a novel pathway to unravel the complex assembly histories of galaxies.

\end{abstract}

\keywords{\uat{Galaxies}{573} --- \uat{Early-type galaxies}{429} --- \uat{Galaxy structure}{622} --- \uat{Galaxy stellar halos}{598} --- \uat{Galaxy environments}{2029} --- \uat{Galaxy mergers}{608} --- \uat{Optical astronomy}{1776}}


\section{Introduction} \label{sec:intro}
Early-type galaxies (ETGs), traditionally classified as elliptical and lenticular systems, are characterized by smooth, spheroidal light distributions and predominantly old stellar populations \cite[][]{1976ApJ...204..668F, 1989ARA&A..27..235K, 2005ApJ...621..673T, 2006ApJ...647L..99D}. This trend is observed consistently across various environments, including even isolated regions \cite[][]{2004MNRAS.354..851R, 2005MNRAS.360..693R, 2007MNRAS.377.1772R, 2016A&A...588A..79L}. Because they are often found in dense environments, such as brightest cluster galaxies in galaxy clusters \cite[][]{1980ApJ...236..351D, 1984ApJ...281...95P, 2003MNRAS.346..601G, 2015MNRAS.451.3427H}, ETGs are generally considered the final stage in galaxy evolution. 

The formation and mass growth of ETGs are believed to be closely tied to galaxy mergers. Major mergers can significantly alter galaxy morphology and kinematics, transforming galaxies of any morphology into spheroids \cite[][]{1977egsp.conf..401T, 1992ApJ...399..462B}, whereas minor mergers contribute to the gradual buildup of the stellar envelope \cite[][]{2009ApJ...699L.178N, 2010ApJ...709.1018V}. Indeed, many observational and theoretical studies consistently point to mergers as the primary driver of morphological change and mass growth in ETGs \cite[e.g.,][]{1996MNRAS.283.1361B, 2004ApJ...616..192C, 2006MNRAS.366..499D, 2013MNRAS.433.2986W}.

Building on the distinct roles of major and minor mergers, a widely accepted scenario is the two-phase growth model: an early dissipative phase of rapid in-situ star formation, followed by a slower phase of ex-situ mass assembly through dry minor mergers \cite[][]{2010ApJ...725.2312O, 2010ApJ...709.1018V}. Under this framework, outer structures---comprising various tidal features and stellar halos\footnote{Technically, stellar halos can be divided into inner and outer components based on their structural properties \cite[see][]{2010MNRAS.406..744C, 2013ApJ...763..113D, 2014MNRAS.444..237P} and stellar populations \cite[see][]{2007Natur.450.1020C, 2012A&A...538A..21S}, which reflect differences in assembly epochs and detailed formation mechanisms. However, we used the term comprehensively in this study.}---serve as crucial imprints of the latter phase. 

Specifically, tidal features are visually compelling tracers of ``recent'' interactions, as their detectability and survival depend on orbital parameters and the surrounding environment \cite[][]{2004IAUS..217..390M, 2008ApJ...689..936J, 2021MNRAS.506.5030M}. In contrast, stellar halos encode the ``cumulative'' assembly history over longer timescales, as they grow by accreting disrupted tidal features \cite[][]{2005ApJ...635..931B, 2013MNRAS.434.3348C, 2016MNRAS.458.2371R}. Therefore, an integrated analysis of both components provides a powerful tool for bridging the observational gap between recent and ancient merging signatures, enabling a comprehensive study of galaxy evolution. 

However, due to the extreme faintness of stellar halos, many observational studies to date have focused on the presence or absence of tidal features and their correlations with other galaxy properties, with particular attention to the role of recent interactions. For instance, \cite{2018ApJ...857..144H} found that galaxies with tidal features tend to be gas-rich. Considering only gas-poor galaxies, those with tidal features are likely to reside in more massive halos and be closer to their nearest neighbors. In addition, \cite{2024MNRAS.529..810R} reported that massive ETGs with tidal features tend to be younger, more massive, and rotate more slowly \cite[see also][]{2023ApJ...946...41Y, 2024ApJ...965..158Y}. 

More recently, \cite{2024ApJ...974..299Y} reported that massive ETGs in low-density environments exhibit tidal features up to approximately three times more often than those in high-density environments, highlighting the role of environmental effects in the detectability of tidal debris. From this finding, we pose a fundamental question: Does this environmental preference merely reflect enhanced debris survivability, or is it a sign of an inherently more active mass assembly history? If the latter is true, the presence of tidal features should serve as a diagnostic indicator of an inherently distinct mass-assembly history, which would manifest as systematic variations in stellar halo structures. 

In this study, we aim to answer this question by comparing radial surface-brightness profiles of massive ETGs with and without tidal features. Rather than focusing on the impact of individual merger events, we investigate whether systematic structural differences emerge in the galaxy outskirts---indicative of a more developed or deficient stellar halo---depending on the presence of tidal features. Furthermore, given that the detection rate of tidal features is strongly coupled to the environment, we scrutinize our analysis by binning the sample by environment and stellar mass to isolate the underlying physical drivers. 

\begin{figure}[t!]
\includegraphics[width=\linewidth]{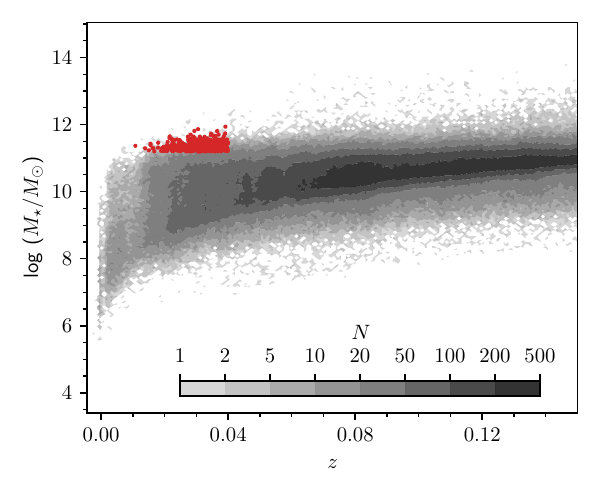}
\caption{Distribution of galaxies in the stellar-mass vs. redshift plane. Grayscale contours show the density of the parent sample from the NSA catalog, with the number of galaxies per bin indicated by the inset colorbar. Red points represent the final sample selected for this study. 
\label{fig:f1}}
\end{figure}

This paper is organized as follows: Section~\ref{sec:data} briefly introduces the dataset and its selection criteria. Section~\ref{sec:method} describes the procedure for measuring the surface-brightness profiles of sampled ETGs. Section~\ref{sec:result} presents results from comparative analyses of their structural properties, and Section~\ref{sec:disc} provides discussion and validation using simulation data. Finally, Section~\ref{sec:sum} summarizes the overall findings of this study. 

\begin{figure*}[t!]
\includegraphics[width=\linewidth]{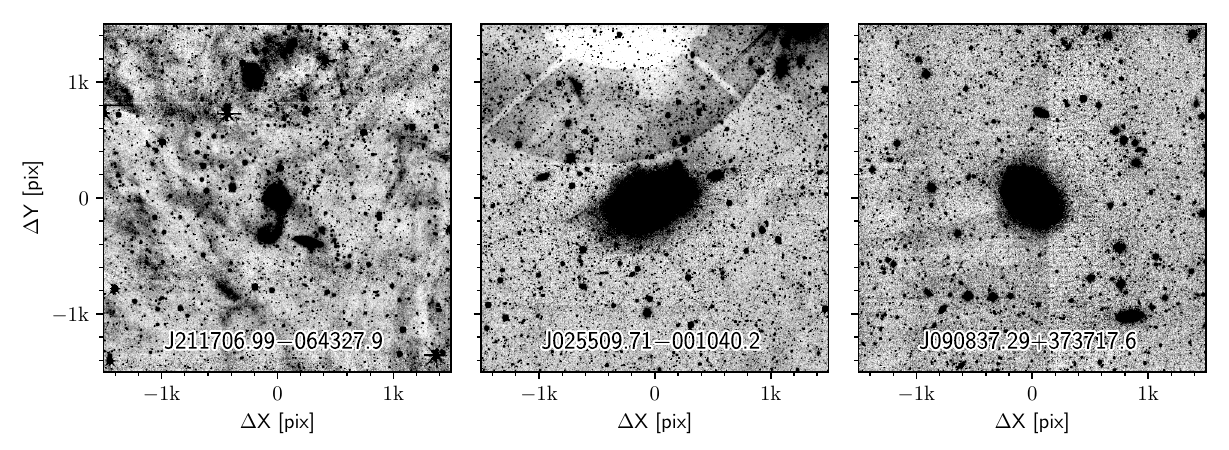}
\caption{Examples of lower-quality data. From left to right, each panel shows contamination from Galactic cirrus, ghost features from a bright star, and artifacts at sensor chip boundaries.
\label{fig:f2}}
\end{figure*}

\section{Data} \label{sec:data}
We used a sample of massive ETGs from \cite{2024ApJ...974..299Y}, which provides a comprehensive catalog of tidal feature classifications and environmental density metrics. Because this catalog was constructed using semi-uniform imaging data and consistent classification criteria, it provides a solid foundation for our structural analysis. The following provides a brief overview of the sample selection process. 

The parent sample comprises approximately 640,000 galaxies from the NASA-Sloan Atlas (NSA) catalog.\footnote{\url{https://www.sdss4.org/dr17/manga/manga-target-selection/nsa/}} The target galaxies were selected primarily within a redshift range of $0.01 < z < 0.04$ and with stellar masses $\mathrm{log}\ (M_\star/M_\odot) > 11.2$, ensuring a sample of massive systems that have likely experienced sufficient mergers to develop extended outer structures \cite[see][]{2017ApJ...834...73Y}. Additionally, only galaxies with reliable stellar mass estimates and high spectroscopic completeness for both target and neighboring galaxies were included. Subsequent visual inspection---utilizing high-resolution color images from the Sloan Digital Sky Survey \cite[SDSS;][]{2000AJ....120.1579Y} and the DESI Legacy Imaging Survey \cite[DLIS;][]{2019AJ....157..168D}---was performed to confirm galaxy morphology, yielding a final sample of 436 ETGs. The sample distributions in the stellar mass-redshift plane are illustrated in Figure~\ref{fig:f1}. 

The identification of tidal features was performed using deep $g$- and $r$-band images from the DLIS Data Release 10.\footnote{\url{https://www.legacysurvey.org}} Specifically, the authors in \cite{2024ApJ...974..299Y} initially identified and classified the tidal features through visual inspection, and we reconfirmed those results. Tidal features were categorized into three primary types: tidal tails---thick, elongated structures originating from host galaxies; stellar streams---narrow, elongated structures likely associated with tidally disrupted companions; and shells---concentric, arc-shaped structures with sharp edges, which often appear in combination. As a result, approximately 45\% (195/436) of the sample exhibit tidal features---regardless of shape or prominence---indicating that nearly half of the massive ETGs may have experienced recent merger events. 

\begin{figure*}[t!]
\centering
\includegraphics[width=\linewidth]{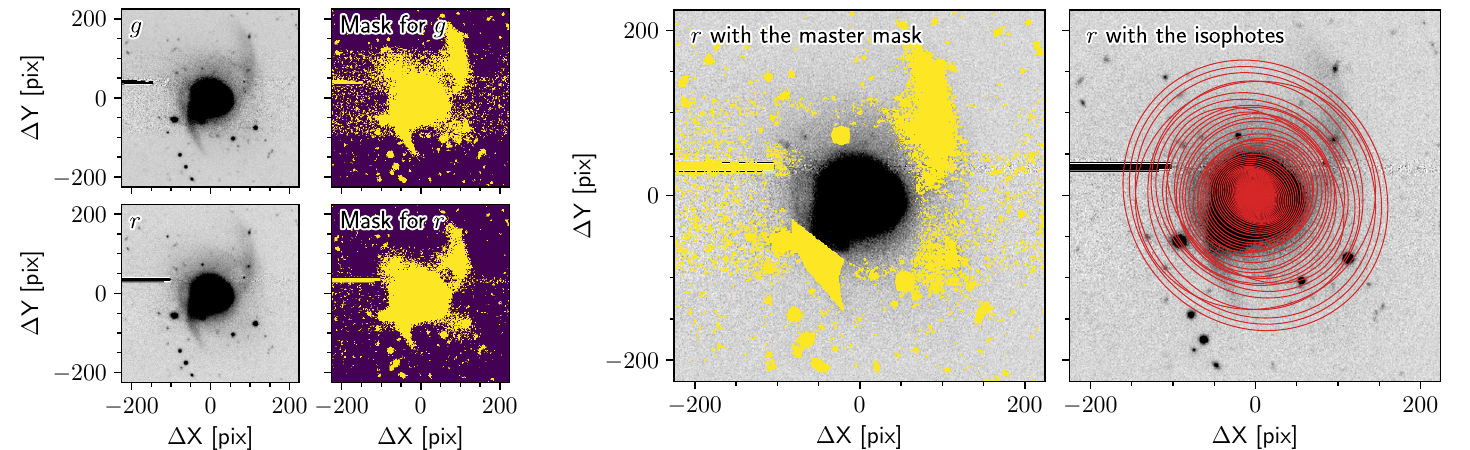}
\caption{Demonstration of the masking and isophote-fitting procedure for the galaxy J094956.38$-$001352.8. The left-hand $2\times 2$ grid shows the original $g$- and $r$-band images, along with their respective initial masks. The two right-hand panels show the $r$-band image with the master mask and the isophote-fitting result overlaid. 
\label{fig:f3}}
\end{figure*}

To extract radial surface-brightness profiles of the sampled galaxies via isophote fitting, we utilized $g$- and $r$-band cutout images from DLIS. These images were generated on a $3\mathrm{k}\times 3\mathrm{k}$ pixel array with a pixel scale of 0\farcs262. Across the images, the $3\sigma$ surface-brightness limit at an angular scale of $10\arcsec\times10\arcsec$ is approximately $28.5\pm0.3$~mag~arcsec$^{-2}$. This photometric dpth is comparable to that of the SDSS Stripe 82 dataset \cite[see][]{2014ApJS..213...12J, 2016MNRAS.456.1359F}; notably, both surveys have been extensively used to study the impact of mergers on galaxy evolution, as they provide the requisite sensitivity to probe the low-surface-brightness (LSB) regime \cite[e.g.,][]{2020ApJ...905..154Y, 2022ApJ...925..168Y, 2023ApJ...946...41Y, 2024ApJ...965..158Y}. 

Although DLIS provides an extensive and valuable dataset, its observational strategy and data processing are not always optimized for LSB studies. Specifically, wide-field ground-based imaging is inherently susceptible to observational challenges, such as localized artifacts and large-scale systematic backgrounds.\footnote{For instance, the DLIS framework sometimes suffers from residual sky patterns caused by the flat-fielding process; see \url{https://www.legacysurvey.org/dr9/sky/} for technical details.} For this reason, we conducted a visual inspection of each cutout image before analysis to identify lower-quality data. We found that approximately 5\% of the images exhibited notable defects, including sky fluctuations from Galactic cirrus, ghost features from bright stars, and artifacts at sensor chip boundaries (Figure~\ref{fig:f2}). 

Such artifacts can contaminate the underlying stellar light and reduce the reliability of radial profiles in the LSB regime.\footnote{Although some of these issues may be partially mitigated through machine-learning-based techniques \cite[see][]{2021A&C....3500469T}, complete removal of artifacts remains fundamentally challenging in the context of deep imaging.} To enhance the purity of radial profile extraction, we excluded them from the subsequent analysis, yielding 184 and 229 galaxies with and without tidal features, respectively. Notably, because their proportions are very small and our analysis relies on the statistical robustness of the median-stacking method, this sampling choice has a negligible effect on the main results. 

\section{Method} \label{sec:method}
\subsection{Object Masking} \label{sec:method:objmask}
Following the initial data quality assessment, it is essential to isolate the target galaxies from other light sources in the images. Even in high-quality images, unmasked foreground stars and background galaxies can affect the surface-brightness profiles, particularly in the faint outskirts of galaxies. To accurately extract the target's radial profile, we generated dedicated mask frames using a hybrid approach that combines automated detection with careful manual refinement, as follows:
\begin{enumerate}
\item{Initial Mask Generation: We produced segmentation maps using \texttt{SExtractor} \cite[][]{1996A&AS..117..393B}. The detection threshold was set to $1\sigma$ above the sky background level to mask even the faintest light sources.}
\item{Target Recovery: Because the target galaxy is inevitably masked in the segmentation maps, we identified the corresponding segments and removed them from the initial masks to expose the target for analysis.}
\item{Master Mask Generation: The masks for the $g$- and $r$-band images were combined into a single master mask frame. This ensures that all contaminating sources detected in either band are excluded, thereby accounting for color-dependent visibility.}
\item{Targeted Manual Refinement: To eliminate potential contaminants that were incompletely masked or accidentally exposed during target recovery, we conducted a comprehensive visual inspection of all object-masked images and applied custom polygon masks to explicitly cover residual tidal features, overlaid contaminants, and optical artifacts.}
\end{enumerate}

The left-hand $2\times2$ grid in Figure~\ref{fig:f3} displays zoomed-in images and initial masks for both bands. Notably, data quality differs between the bands: a stripe-like noise pattern---likely caused by the sensor chip gap or stray light from nearby stars---crosses the target galaxy in the $g$-band image, producing slight differences in the initial masks derived from each band. Additionally, the saturation trail on the left side of each image appears differently across the bands. 

The two right-hand panels in Figure~\ref{fig:f3} show the master mask and the isophote-fitting results. This galaxy clearly illustrates how our hybrid masking procedure handles different tidal structures. The prominent, nearly detached tidal tail extending to the north was identified as an independent segment by the automated detection algorithm and thus remained cleanly masked after the target recovery phase. In contrast, the thinner and/or shorter features extending to the south were initially blended with the galaxy’s main body and inadvertently unmasked during target recovery. To prevent this structure from distorting the radial profile, we applied a custom polygon mask to fully cover the exposed debris. As evidenced by the smooth, undistorted isophotes in the rightmost panel, this targeted refinement ensures that the isophote fitting robustly captures the underlying diffuse envelope without being perturbed by asymmetric tidal features. 

Indeed, we verified the robustness of our results to the masking scheme. Because isophotes are measured from fixed geometric centers with azimuthal 3$\sigma$ clipping, minor residual features along individual isophotes are naturally rejected. To quantitatively evaluate any potential systematic uncertainty introduced by masking choices, we compared radial profiles extracted before and after applying our targeted manual refinements. Across the sample, profile variations remained minor ($\lesssim$0.05~dex) and localized, while characteristic radial scales, such as the half-light radius, shifted by less than 5\%. Crucially, these minor perturbations cannot systematically alter the global radial profile, confirming that our final comparative results remain robust to masking choices. 

\subsection{Isophote fit} \label{sec:method:fit}
Building on the resulting masks, we conducted elliptical isophote fitting for the target galaxies using the Python package \texttt{photutils.isophote}. To ensure consistent results across the sample, the fitting was performed based on the following configurations:
\begin{itemize}
\item{Geometric Parameters: The initial isophote parameters were determined by automated optimization. Subsequently, the ellipticity and position angle were allowed to vary freely at each radius to capture structural variations. The central position was held fixed to mitigate contamination from unmasked sources, which could otherwise destabilize the fit at large radii.}
\item{Sampling and Integration: The semimajor axis was sampled logarithmically, with each successive isophote radius changing by a factor of $(1+\texttt{step})$, where \texttt{step} is 0.05. We employed the ``bilinear'' area-integration mode to balance sampling precision and speed. }
\item{Outlier Rejection: To suppress the impact of unmasked sources, we applied $3\sigma$ clipping to data points on each isophote. Furthermore, we relaxed the \texttt{fflag}---the fraction of rejected data points---threshold to 0.3 to prevent premature termination of the fitting process in galaxies with extensive masking due to tidal features. }
\end{itemize}

When the initial fit failed to converge, we adjusted the starting semimajor axis or lowered the \texttt{fflag} threshold to obtain a solution. These adjustments were primarily needed to initiate the process for galaxies and did not introduce artificial features into the resulting radial profiles. 

It is worth noting that the choice of area-integration mode may be critical, particularly in the LSB regime. While the median mode is traditionally more robust against noise, it is significantly more computationally expensive for large samples. To justify our use of the bilinear mode, we cross-validated its performance against the median mode on a subset of our sample. The resulting profiles were nearly identical in shape, with differences observed only in the maximum semimajor axis length. These variations showed no systematic trend, indicating they were random responses to local noise rather than an inherent bias. Consequently, we concluded that our choice is reliable and does not compromise the results of this study. 

\subsection{Normalization and Scaling of Radial Profiles} \label{sec:method:norm}
The primary objective of this study is to compare the azimuthally averaged surface-brightness profiles of galaxies with tidal features (GwT) and those without (GwoT), particularly in their outskirts. However, these apparent profiles are often difficult to compare directly because of variations in redshift, morphology, stellar mass, and stellar population. To ensure a fair comparison and reveal the structural signatures of interest, it is essential to normalize the individual profiles to a common scale. 

\subsubsection{Limitations of the Outermost Radius as a Scaling Reference} \label{sec:method:norm:rmax}
For a sample of massive ETGs, which typically exhibit monotonic structural properties, the maximum isophotal radius ($r_\mathrm{max}$) might naively seem a convenient basis for profile comparison. However, $r_\mathrm{max}$ is fundamentally unsuitable as a scaling metric for profile normalization for two critical reasons. 

\begin{figure}[t!]
\includegraphics[width=\linewidth]{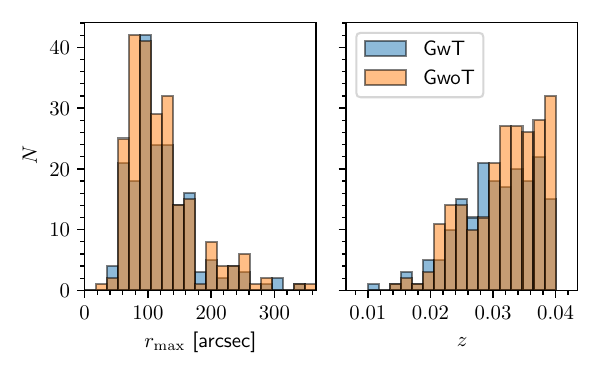}
\caption{Histograms of the maximum isophotal radius (left) and redshift (right) for the GwT and GwoT samples. According to the Anderson--Darling test, the redshift distributions of the two samples show a marginally distinct trend ($p=0.07$).
\label{fig:f4}}
\end{figure}

First, in angular units, $r_\mathrm{max}$ is inherently redshift-dependent and therefore fails to provide a consistent scaling across a sample with varying distances. In our case, the GwoT sample is slightly biased toward higher redshifts than the GwT sample (Figure~\ref{fig:f4}), which would potentially introduce an artificial offset in any angular-based comparison. Second, and more importantly, converting these measurements to physical units (kpc) does not yield a more reliable normalization. This is because the measured angular extent is not an intrinsic physical boundary but is determined by photometric depth and the accuracy of local sky subtraction. It is also affected by the specific setup of the isophote fitting, such as the chosen \texttt{step} size or stopping criteria.\footnote{Given the \texttt{step} size of 0.05 in the isophote fitting of this study, some of $r_\mathrm{max}$ may be underestimated by up to 5\%.} 
 
To enable a physically meaningful comparison, the radial profiles must be normalized using a physically motivated basis. At this point, it would be particularly beneficial if such a scale could also distinguish between inner and outer structures, thereby facilitating a more focused analysis of the stellar halos of interest in this study. 

\subsubsection{Determination of the Normalization Scale} \label{sec:method:norm:r50}
Defining a universal boundary for stellar halos is challenging, as the transition radius where ex-situ stellar populations dominate can vary from 10 to 50~kpc \cite[e.g.,][]{2013MNRAS.432.3391T,2014MNRAS.443.1433D,2018ApJ...869...12S}. Instead, various alternative metrics have been proposed in the literature, including multiples of the disk scale length \cite[e.g.,][]{2018ApJ...869...12S}, half-mass radius \cite[e.g.,][]{2014MNRAS.444..237P,2016ApJ...830...62M,2022ApJ...932...44G}, half-light radius \cite[e.g.,][]{2023MNRAS.523..123S,2023MNRAS.525.2875S,2023MNRAS.519.1349W,2024MNRAS.527.4209J,2024AJ....167...57T,2025A&A...699A..93T}, or a fixed physical radius \cite[e.g.,][]{2020MNRAS.495.4570M}. In this study, we adopt the half-light radius ($r_{50}$) as our standard scaling metric for profile normalization. 

\begin{figure}[t!]
\includegraphics[width=\linewidth]{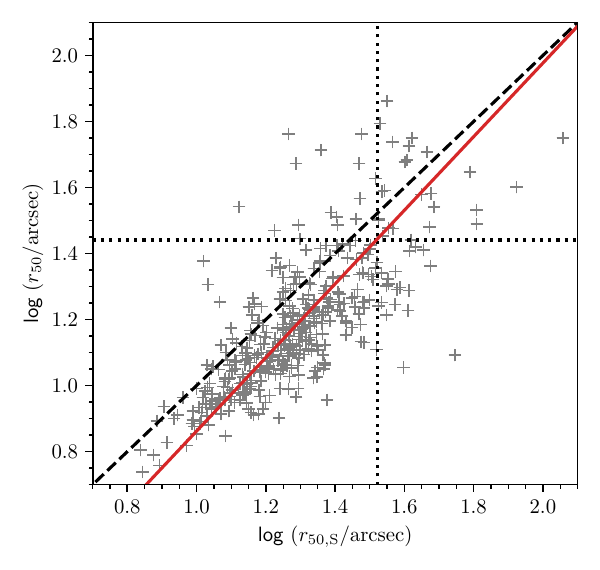}
\caption{Comparison between our nonparametric $r_{50}$ and the Sérsic-based $r_\mathrm{50,S}$ from the NSA catalog. The dashed black line indicates a one-to-one relation, while the solid red line shows the result of an orthogonal distance regression fit. The dotted black lines denote the 90th percentiles of the distribution.
\label{fig:f5}}
\end{figure}

Although the NSA catalog provides the half-light radius derived from two-dimensional Sérsic model fits ($r_\mathrm{50,S}$), such parametric measurements are often sensitive to the fitting scheme: because the NSA data are shallower than DLIS data, they may be strongly influenced by the bright inner regions and artificially inflated by analytical extrapolation into the model's infinite tail. For this reason, we chose to compute independent empirical $r_{50}$ values directly from the observed curves of growth, thereby avoiding potential biases arising from varying galaxy geometries and orientations. 

Specifically, we assumed circular symmetry for all targets, treating each azimuthally averaged surface-brightness profile as a series of concentric circular isophotes. From these, we constructed a one-dimensional cumulative light profile by integrating the flux radially over the two-dimensional plane. The radius at which the cumulative flux reaches 50\% of the total integrated light was defined as $r_{50}$. This circularized, nonparametric approach standardizes the radial scaling across galaxies with diverse inclinations and axis ratios, effectively mitigating projection-dependent systematics in our comparative analysis. 

\begin{figure*}[t!]
\includegraphics[width=\linewidth]{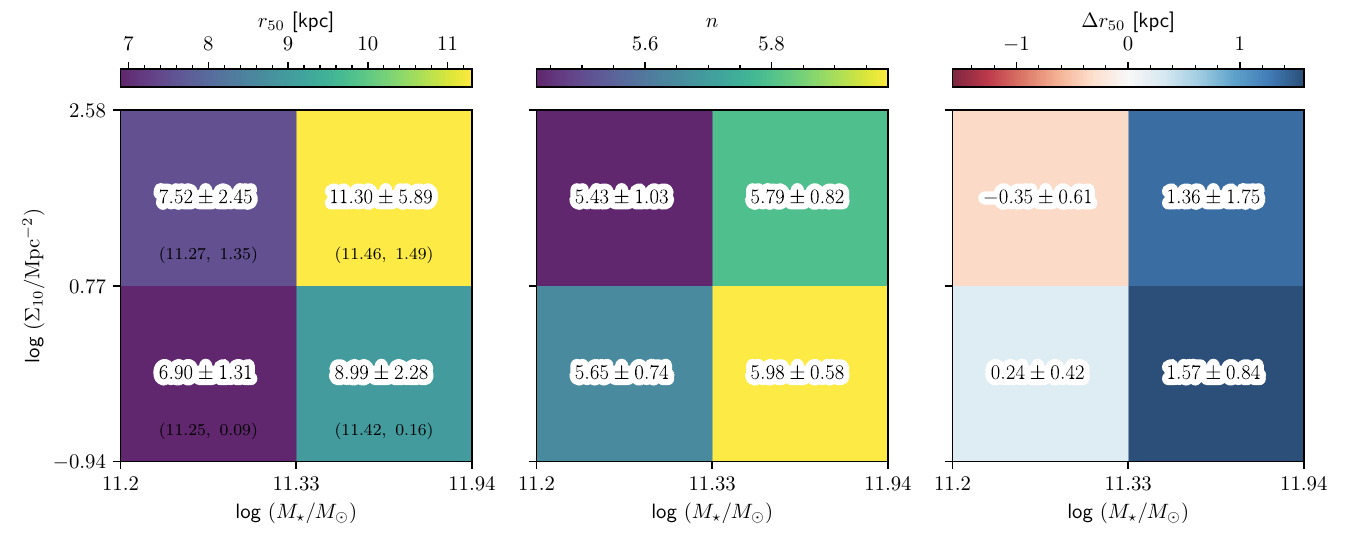}
\caption{Structural scaling relations across the environmental density vs. stellar-mass grid. The left and middle panels show the median values of $r_{50}$ and $n$ for the entire sample, while the right panel shows the residuals $\Delta r_{50}$ between GwT and GwoT. The numbers overlaid in each grid cell indicate the associated uncertainties: standard deviations for $r_{50}$ and $n$, and the quadrature sum of the bootstrap standard errors of the two median values of $\Delta r_{50}$ for $\Delta r_{50}$. In the left panel, the median stellar mass and environmental density for each bin are explicitly quoted in parentheses at the bottom of each cell.
\label{fig:f6}}
\end{figure*}

Figure~\ref{fig:f5} compares our computed $r_{50}$ with the Sérsic-based $r_\mathrm{50,S}$. A systematic offset is evident, with our $r_{50}$ measurements roughly 0--20\% smaller than $r_\mathrm{50,S}$. This discrepancy reflects methodological differences: Sérsic-based measurements incorporate analytical extrapolation beyond $r_\mathrm{max}$, whereas our integration is strictly constrained to $r_\mathrm{max}$. This offset does not indicate a drawback in either approach; rather, it demonstrates that our measurements provide a reliable and consistent basis for internal comparisons across the sample. 

Finally, to focus our analysis on the outer regions most likely shaped by stellar halos, we defined a transition radius, $r_t\equiv 5\times r_{50}$, as our final scaling metric for profile normalization. This choice is strongly supported by two physical indicators. First, the median physical radius at $5\times r_{50}$ is approximately 40~kpc, which lies at the upper end of the transition-radius range noted at the beginning of this section---a reasonable expectation for massive ETGs. Second, the median surface brightness at this radius for our sample is $\sim$26~mag~arcsec$^{-2}$, marginally meeting the threshold at which ex-situ stellar halos become dominant \cite[see][]{2011ApJ...739...20C, 2012arXiv1204.3082B}. Given this agreement, our choice of transition radius provides a physically consistent and robust anchor for our structural analysis. 

\section{Results} \label{sec:result}
Having established a consistent normalization framework based on the half-light radius as our primary scaling metric, we now examine structural differences between the GwT and GwoT samples by comparing their radial surface-brightness profiles. 

Before starting the comparative analysis, we conducted a preliminary consistency check to assess the reliability of our $r_{50}$ measurements. We compared the $r_{50}$ values derived independently from the $g$- and $r$-band images and found excellent agreement, with a residual standard deviation of only 0.74 arcsec. Given this high degree of consistency, we adopted the $r$-band profiles as the primary basis for all subsequent analyses, as they trace the underlying stellar components---likely old populations---more effectively and ensure methodological uniformity. 

\subsection{Systematic Trends of $r_{50}$} \label{sec:result:r50}
We first examined the $r_{50}$ distributions of the GwT and GwoT samples to establish a structural baseline. Although $r_{50}$ lies in the inner regions by definition of our transition radius, it may serve as an indirect indicator of the global light distribution and provide necessary context for interpreting the subsequent analyses of radial surface-brightness profiles. 

Given that $r_{50}$ is strongly correlated with both the local environment and stellar mass, we divided our sample into a $2\times2$ grid to disentangle the relative contributions of these two factors. Following \cite{2024ApJ...974..299Y}, we adopted the surface number density ($\Sigma_{10}\equiv10/\pi r_{10}^2$) as our environmental metric, which measures local density using the projected area defined by the distance to the tenth-closest neighbor ($r_{10}$). Each grid axis was binned independently using its median, and this straightforward division yields a variable galaxy count per grid cell---either 83 or 123.\footnote{The density bins roughly separate environments that are denser or less dense than the group-scale environments, in terms of physical scale.} 

For reference, the median stellar mass and environmental density for each of the four bins are explicitly noted in Figure~\ref{fig:f6}. The residual correlations between stellar mass and environmental density across the bins are minimal, confirming that our discrete binning effectively decouples the two parameters without introducing notable sampling artifacts. This ensures that the structural variations discussed in this and subsequent sections reflect genuine physical effects. 

The left panel of Figure~\ref{fig:f6} shows the median $r_{50}$ values across the density-mass grid. We found that $r_{50}$ increases monotonically with both environmental density and stellar mass. While these trends probably reflect a combination of evolutionary effects, including quenching-driven changes in the inner profile, we regard them primarily as an expansion of galaxies' physical scale. In other words, galaxies in denser environments and with higher stellar masses tend to have more extended light distributions. 

As a supplementary check, we examined the Sérsic index ($n$) from the NSA catalog to assess whether the observed trend in $r_{50}$ is accompanied by changes in central concentration. As shown in the middle panel of Figure~\ref{fig:f6}, $n$ increases with stellar mass but tends to decrease with environmental density; however, these trends are highly tentative given the large scatter and marginal differences between the grid cells. Consequently, these variations are not definitive enough to confirm any systematic change in profile shape, suggesting that, for the galaxies in our sample, structural expansion is a more plausible cause of the growth in $r_{50}$. Although inferential, such divergent trends in $r_{50}$ and $n$ may hint at distinct pathways of galaxy growth: mass-driven growth associated with concentration versus environment-driven build-up of extended envelopes. 

Finally, we examined the difference in $r_{50}$ between the GwT and GwoT samples. The right panel of Figure~\ref{fig:f6} shows the residuals ($\Delta r_{50}$), obtained by subtracting the median $r_{50}$ of GwoT from that of GwT. We found that the residuals are minimal in the lower-mass bins---irrespective of sign---but become pronounced in the higher-mass bins, where GwT tends to be larger than GwoT. Specifically, the relative offset ($\Delta r_{50}/r_{50}$) increases from $\sim$4\% to $\sim$15\% with stellar mass. The environmental effect acts slightly differently across the mass regimes: As $\Sigma_{10}$ increases, $\Delta r_{50}$ decreases moderately in the lower-mass bins but only subtly in the higher-mass bins. 

Importantly, we confirmed that the GwT and GwoT subsamples exhibit nearly identical distributions of intrinsic properties. Specifically, the stellar mass and Sérsic index distributions are statistically indistinguishable between the two subsamples, with Anderson--Darling tests yielding $p$-values greater than 0.25. Only the median $u-r$ color of GwoT is slightly redder by $\sim$0.02 mag, but this difference is well within the typical $u$-band magnitude uncertainty. This demonstrates that the observed differences in $r_{50}$ are not driven by underlying stellar-population disparities. Taken together, these divergent behaviors---whether consistent or reversed depending on the case---imply that mass and environment may drive structural evolution through distinct physical channels, suggesting that a grid-based approach can also be effective for investigating structural signatures in outer surface-brightness profiles. 

\subsection{Comparison of Radial Surface-Brightness Profiles} \label{sec:result:profile}
To investigate whether GwT and GwoT exhibit different light distributions in their outskirts, we compared their azimuthally averaged surface-brightness profiles. As described in Section~\ref{sec:method:norm}, we normalized each profile at the transition radius $r_t = 5\times r_{50}$. Galaxies with $r_\mathrm{max}$ smaller than $r_t$ were excluded from the normalization, yielding a final sample of 172 GwT and 209 GwoT for comparison. 

\begin{figure}[t!]
\includegraphics[width=\linewidth]{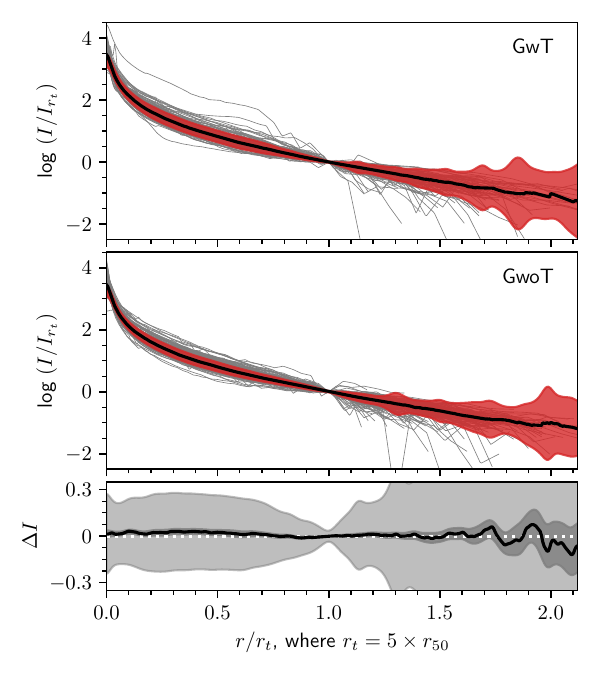}
\caption{One-dimensional radial surface-brightness profiles of GwT (top) and GwoT (middle), normalized to $r_t=5\times r_{50}$. Individual profiles are shown as thin gray lines, with median-stacked results and their uncertainties overlaid in black lines and red shaded regions, respectively. The bottom panel shows the residual, $\Delta I = \log\, I_\mathrm{GwT} - \log\, I_\mathrm{GwoT}$, calculated from the two median-stacked profiles. The light-gray and dark-gray shaded regions represent the propagated uncertainty (quadrature sum) and the bootstrap standard error of the median difference, respectively. 
\label{fig:f7}}
\end{figure}

It is worth noting that the normalization was applied to both the radial scale ($r/r_t$) and the surface brightness ($I/I_{r_t}$). This approach effectively mitigates luminosity bias and decouples the inner and outer structures, allowing us to focus on the relative shapes of the stellar halo profiles. Consequently, the intrinsic scatter among individual profiles and the associated statistical uncertainties in the profiles naturally converge to zero at the transition point $r/r_t=1$. 

\subsubsection{Comparison for the Entire Sample} \label{sec:result:profile:entire}
Figure~\ref{fig:f7} shows the radial surface-brightness profiles for the GwT and GwoT samples, including both individual profiles and their ensemble profiles. Because our primary interest is the ensemble profile, we briefly note two apparent features in the individual profiles: some GwT profiles deviate in their inner regions, likely due to interaction-driven disturbances, and both samples occasionally show truncations in their outer regions. 

To derive the ensemble profiles, we median-stacked the surface-brightness profiles at each normalized radius. The uncertainty estimates for these stacked profiles combine the mean isophote-fitting errors with the profile-to-profile dispersion (standard deviation) at each radial bin. To account for the reduced statistical power in the outskirts---caused by varying normalized $r_\mathrm{max}$---we scaled the errors by $\sqrt{N_t/N_i}$, where $N_t$ is the total number of galaxies and $N_i$ is the number of galaxies contributing at a given radius. Finally, to ensure the reliability of our median estimates, we restricted our analysis to radial bins with contributions from at least five galaxies. 

Here, it should be noted that investigating the diffuse outskirts of galaxies is inherently limited by the photometric depth of the imaging data. Therefore, the large uncertainties in these LSB regimes make it challenging to draw definitive physical conclusions. Nevertheless, we emphasize that relative differences between the two samples can still provide meaningful insights if they manifest as consistent, systematic trends. 

Using ensemble profiles, we examined structural differences between the two samples. As shown in the bottom panel of Figure~\ref{fig:f7}, the residual ($\Delta I$)---calculated by subtracting the GwoT profile from the GwT profile---remains remarkably near zero within the uncertainties across all radial bins. 

\begin{figure}[t!]
\includegraphics[width=\linewidth]{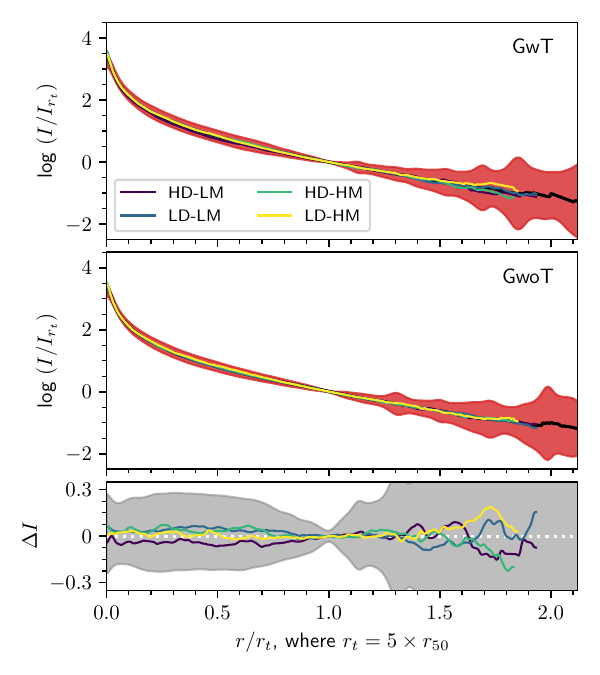}
\caption{Same as Figure~\ref{fig:f7}, but partitioned into the $2\times2$ density-mass grid. The color-coded lines correspond to the four regimes (HD-LM, LD-LM, HD-HM, and LD-HM) and are arranged according to the grid layout. 
\label{fig:f8}}
\end{figure}

This global similarity suggests that the presence or absence of detectable tidal features does not necessarily reflect distinct mass-assembly histories that would affect the development of stellar halos. Alternatively, subtle structural differences may be diluted by intrinsic and extrinsic factors. Specifically, as described in Section~\ref{sec:result:r50}, local environment and stellar mass are likely the primary drivers of structural variation. Therefore, a more controlled analysis using grid-based separation is essential. 

\subsubsection{Comparison across the Density-Mass Grid} \label{sec:result:profile:grid}
To disentangle the competing impacts of the local environment and stellar mass on the observed structural signatures, we repeated the profile comparison using the $2\times 2$ density-mass grid defined in Section~\ref{sec:result:r50}. Figure~\ref{fig:f8} presents the radial surface-brightness profiles for the resulting four regimes: higher density \& lower mass (HD-LM), lower density \& lower mass (LD-LM), higher density \& higher mass (HD-HM), and lower density \& higher mass (LD-HM). To maintain adequate radial depth within these smaller subsamples, we slightly relaxed the minimum galaxy count for median-stacking from five to three per bin. 

\begin{figure}[t!]
\includegraphics[width=\linewidth]{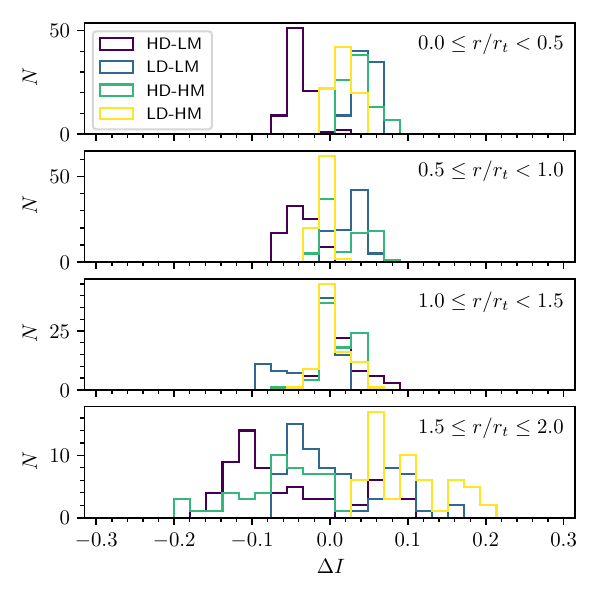}
\caption{Distributions of the residuals across four successive radial bins, from the inner regions (top) to the outskirts (bottom). Each histogram is color-coded by density-mass regime (HD-LM, LD-LM, HD-HM, and LD-HM). 
\label{fig:f9}}
\end{figure}

Across all regimes, the ensemble profiles remain broadly consistent for $r/r_t < 1.5$, showing no prominent trends linked to either environment or stellar mass. However, the residuals begin to emerge beyond $r/r_t \sim 1.5$. Specifically, the LD-HM regime exhibits a marginal positive residual, whereas the HD-LM regime shows a slight negative residual. This trend is further clarified by the $\Delta I$ histograms in uniform radial bins (Figure~\ref{fig:f9}). While these residuals remain centered near zero at $r/r_t < 1.5$, a noticeable shift---positive for the LD-HM regime and negative for the HD-LM regime---occurs at $1.5 \le r/r_t \le 2.0$. According to a two-sided Wilcoxon signed-rank test against the null baseline ($\Delta I = 0$), the outskirts exhibit statistically significant shifts: the LD-HM regime shows a tentative excess (median $\Delta I = +0.10$, $p$-value = $7.6\times10^{-11}$), whereas the HD-LM regime shows a distinct deficit (median $\Delta I = -0.08$, $p$-value = $8.3\times10^{-6}$). 

We emphasize that interpreting these signatures requires strong caution. Because the galaxy sample is partitioned into four bins, the reduced number of galaxies per cell amplifies the intrinsic profile dispersion, leading to large uncertainties in the outskirts. Furthermore, the observed profile differences appear predominantly within a localized radial range and are sensitive to the chosen radial scaling scheme. Rather than claiming definitive, globally systematic structural differences, we interpret these patterns as suggestive signatures of differential stellar halo assembly and environmental disruption that warrant verification with larger statistical samples and deeper imaging datasets. 

\section{Discussion} \label{sec:disc}
\subsection{Interplay of Environment and Stellar Mass in Shaping Stellar Halos} \label{sec:disc:origin}
Our primary hypothesis is that the presence of tidal features serves as a visible proxy for a systematically more active and sustained mass-assembly history. However, our grid-based analysis revealed that the structural signal in the galaxy outskirts---the development of the stellar halo---is not uniform. Instead, the distinction between GwT and GwoT appears to be modulated by two competing physical factors: mass-driven growth potential and environment-driven preservation efficiency. 

Although we divided our sample by stellar mass for relative comparison, all host galaxies in this study are fundamentally massive ETGs, and the median stellar mass difference between the LM and HM bins is modest ($\sim$0.2~dex). Given these modest mass contrasts, the LD-HM regime is highlighted not as a definitive benchmark for all massive systems, but rather as an illustrative physical case in which competing external effects are minimized. Theoretically, massive hosts provide substantial accretion potential, while lower-density environments minimize environmental perturbations (e.g., severe tidal stripping, harassment, or dilution by intracluster light). Therefore, the LD-HM regime serves as a favorable case study for assessing whether cumulative accretion leaves observable imprints in the outer envelope. Under this interpretation, the currently observed tidal features in massive, field/group galaxies may reflect a more sustained accretion history, allowing outer stellar envelopes to remain observable over longer timescales. 

\paragraph{Enhanced Outer Light in the LD-HM Regime} The LD-HM regime exhibits a noticeable excess in outer light for GwT relative to GwoT, which is qualitatively consistent with an assembly-driven scenario. In this regime, massive hosts facilitate significant accretion, while the lower-density environment minimizes severe environmental disruption, such as strong tidal stripping or rapid dispersion. Under this interpretation, the currently observed tidal features in massive, field/group galaxies may be accompanied by a more sustained accretion history, allowing outer stellar envelopes to remain observable over longer timescales. 

\paragraph{Apparent Deficits in the HD-LM Regime} Conversely, in the HD-LM regime, GwT exhibits relatively lower outer surface brightness compared to GwoT. One plausible interpretation is that in dense environments, lower-mass hosts are more susceptible to external perturbations, such as galaxy harassment and cluster tidal fields. \cite[see][]{2001AJ....121..808C, 2012AJ....144..128A, 2014A&A...566A..97J, 2024ApJ...974..299Y}. In such dense regimes, detectable tidal features might reflect an active, transient disruption phase where outer material is readily liberated or stripped into the surrounding halo, temporarily outpacing the cumulative build-up of a bound stellar envelope. 

\paragraph{Intermediate Behaviors in the LD-LM and HD-HM Regimes} The remaining two regimes show minimal differences between GwT and GwoT, pointing to regimes where structural divergence is either intrinsically subtle or counterbalanced. In the LD-LM regime, despite favorable preservation conditions, lower host masses naturally limit the total accreted stellar mass, keeping profile differences marginal. In the HD-HM regime, the competing effects of high accretion capacity and strong environmental perturbations may moderate observable differences in the outer profiles. 

\paragraph{Summary and Alternative Considerations} In summary, while these trends suggest a plausible scenario in which the formation/disruption of tidal features and stellar halos depend on host mass and local density, we caution that other physical mechanisms could also contribute to the observed outer profiles. These include variations in merger orbits, progenitor mass ratios, and the complex growth timescales of stellar envelopes, as well as observational challenges such as tidal debris being veiled or diluted by intracluster light in dense environments \cite[see][]{2026ApJ...996..145C}. Detailed kinematic studies and numerical simulations will be essential to definitively disentangle these competing processes. Nonetheless, our empirical findings highlight that interpreting tidal features and stellar halo growth requires considering the broader environmental and mass context. 

\subsection{Simulation-Based Validation} \label{sec:disc:valid}
The physical scenario proposed in the previous section requires independent verification within a controlled numerical framework. To this end, we leveraged the $N$-body cosmological simulations to examine whether the structural differences identified in our observational work reflect genuine physical signatures of mass assembly rather than artifacts of our measurement or normalization schemes. The simulations provide a unique opportunity to directly correlate the current stellar distribution with the underlying merger history. Specifically, we aim to examine (1) how the presence of tidal features affects mock surface-brightness profiles under controlled conditions, and (2) whether GwT indeed possesses a systematically more active mass-assembly history, as hypothesized.

\subsubsection{Mock Data} \label{sec:disc:valid:data}
We utilized simulation data generated via the galaxy replacement technique \cite[GRT;][]{2022ApJ...925..103C}. GRT is an optimized method for simulating galaxy mass assembly by replacing low-resolution dark matter halos in cosmological volumes with high-resolution galaxy models, thereby enabling efficient tracing of galaxies' merger and accretion histories. This approach is particularly suitable for our study, as it has been validated to reproduce both the stellar mass growth of central galaxies and the distribution of their surrounding diffuse light \cite[see][]{2023ApJ...943..148C, 2024ApJ...969..142C, 2026ApJ...996..145C}. 

To account for observational effects under a conservative baseline, the simulated galaxies were projected at a fiducial distance of 200~Mpc---slightly beyond the maximum distance in our observational sample. This distance was chosen to set a stringent resolution requirement, ensuring that the simulation's mass resolution ($7.9\times10^4~M_\odot$) and gravitational softening length (14.6~pc) remain sufficient to resolve features down to our observational limits.

The stellar mass for each candidate was computed by integrating the stellar particle masses within the virial radius, yielding a parent sample of 70 massive ETGs. Although a subset of these galaxies fell slightly below the $10^{11}~M_\odot$ threshold under an SDSS-like surface-brightness cut, we retained them in the analysis to preserve sample completeness, as they still occupy the high-mass regime of interest. For each of the 70 parent galaxies, we generated two-dimensional particle maps along three orthogonal axes ($xyz$), yielding 210 sample realizations. This approach not only increases the sample size to enhance statistical robustness but also allows us to directly assess how projection effects influence the detectability of tidal debris (see Section \ref{sec:disc:valid:impact}). 

\begin{figure}[t!]
\includegraphics[width=\linewidth]{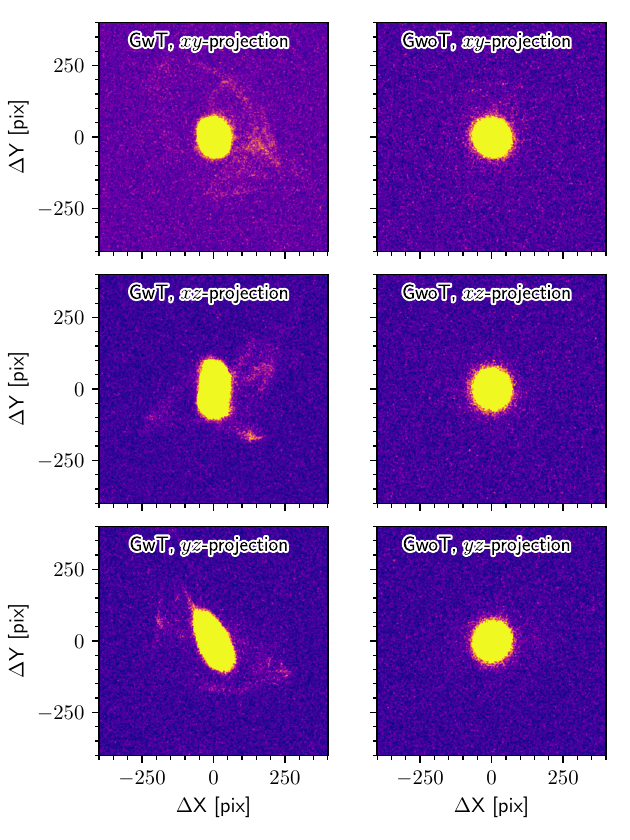}
\caption{Example mock images of galaxies from the GRT simulation. The left and right columns show simulated GwT and GwoT, respectively. For each system, the three rows represent orthogonal projections to illustrate the three-dimensional distribution of stellar material. 
\label{fig:f10}}
\end{figure}

To generate mock images directly comparable to the DLIS $r$-band data, we assigned $r$-band luminosities to individual stellar particles using simple stellar population models based on the spectral evolution models of \cite{2003MNRAS.344.1000B}, with metallicities adopted from the galaxy mass--metallicity relation \cite[][]{2016MNRAS.456.2140M}.\footnote{See details in \cite{2024ApJ...969..142C}} We then integrated the flux within each $0\farcs262 \times 0\farcs262$ spatial pixel, and converted the counts to standard flux units consistent with the DLIS photometric zeropoint. The resulting images were then convolved with an observational point-spread function (PSF; a Moffat kernel with $\mathrm{FWHM} \sim 0\farcs9$)\footnote{\url{https://www.legacysurvey.org/dr9/psf/}} to replicate DLIS images. Crucially, these images were generated exclusively from stellar particles associated with the target galaxies, excluding all other components (e.g., foreground stars and background galaxies). Finally, we added Gaussian background noise calibrated to reproduce the survey's nominal $3\sigma$ surface-brightness limit ($\mu_r\sim 28.5$~mag~arcsec$^{-2}$ on a $10\arcsec\times10\arcsec$ scale). 

After data utilization, the resulting mock images were processed with the same analysis pipeline used for the observational data, including the isophote-fitting procedure and tidal-feature identification. To minimize subjective bias, tidal features were identified through repeated blind visual inspections. Specifically, each realization was evaluated multiple times in a randomized order, and the final classification was determined by the majority consensus of these repeated blind classifications. Through this rigorous procedure, 120 of 210 realizations ($\sim$57\%) were ultimately classified as exhibiting detectable tidal features. Figure~\ref{fig:f10} shows examples of simulated GwT and GwoT projected along three orthogonal axes. 

Although the mock sample is smaller than the observational dataset---nearly half the size---the simulation-based test provides a controlled environment free of celestial contamination and instrumental artifacts. This ensures that any detected structural variations are driven solely by galaxies' properties. Furthermore, by maintaining uniform physical resolution, we effectively isolate intrinsic physical effects from systematics arising from resolution variations with distance, thereby providing a robust test of whether the intensity offset in the outer regions of profiles is a fundamental consequence of ex-situ structural evolution.

\subsubsection{Comparison of Mock Radial Surface-Brightness Profiles} \label{sec:disc:valid:result}
\begin{figure}[t!]
\includegraphics[width=\linewidth]{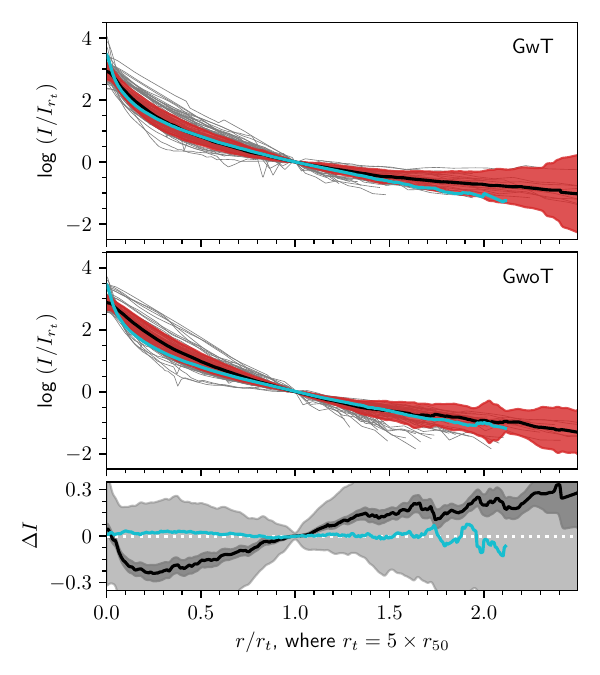}
\caption{Same as Figure~\ref{fig:f7}, but with the simulated galaxies. For visual comparison, the observational ensemble profiles are overlaid as solid cyan lines. 
\label{fig:f11}}
\end{figure}

To explore the physical origins of the observed structural variations in the galaxy outskirts, we compared the radial surface-brightness profiles of the simulated GwT and GwoT, following the same procedure as for the observational data. The results are shown in Figure~\ref{fig:f11}. 

Before comparing the simulated GwT and GwoT profiles, we examined the overall profile characteristics of the simulation relative to the observation. First, in the inner regions, the mock profiles exhibit lower central concentrations than the observational profiles, a deficit that is particularly pronounced in the GwoT sample. This difference is attributed to the dissipationless $N$-body nature of the GRT simulation, which lacks the steep central cores seen in real massive ETGs \cite[see][]{2007ApJ...658...65C, 2007ApJ...658..710N, 2009ApJS..182..216K}. Second, in the outer regions, the simulated GwT hosts exhibit a surface-brightness excess compared to the observational GwT. Third, the mock profiles appear to extend farther in normalized radial units. While one might attribute this to the inherently clean nature of the simulation allowing profile recovery without truncation, this apparent extension is in fact primarily driven by the scaling: the simulated galaxies exhibit an inherently smaller median $r_{50}$ value (by $\sim$30\%) than the observational sample. 

We now shift our attention to the residuals between the simulated GwT and GwoT (bottom panel of Figure~\ref{fig:f11}). The most noticeable feature of the mock residual profile is a distinct bimodal trend: a prominent inner trough followed by a consistent outer excess. This pattern differs markedly from the observational results, suggesting that the simulation may be influenced by additional mechanisms. To pinpoint the specific drivers of this signal and isolate the underlying signatures, we applied a mass-based stratification, similar to our observational framework. 

We caution that the GRT simulation was conducted exclusively in a cluster environment, so the environmental density definition used in our observational analysis is not directly applicable. Although alternative proxies for environmental density---such as a galaxy's infall time into the cluster or a cumulative tidal perturbation index---exist, these measures can introduce unwarranted complexity and interpretive ambiguity. To avoid these issues and ensure a robust comparison, we chose to focus the simulation analysis primarily on the impact of stellar mass. 

\begin{figure}[t!]
\includegraphics[width=\linewidth]{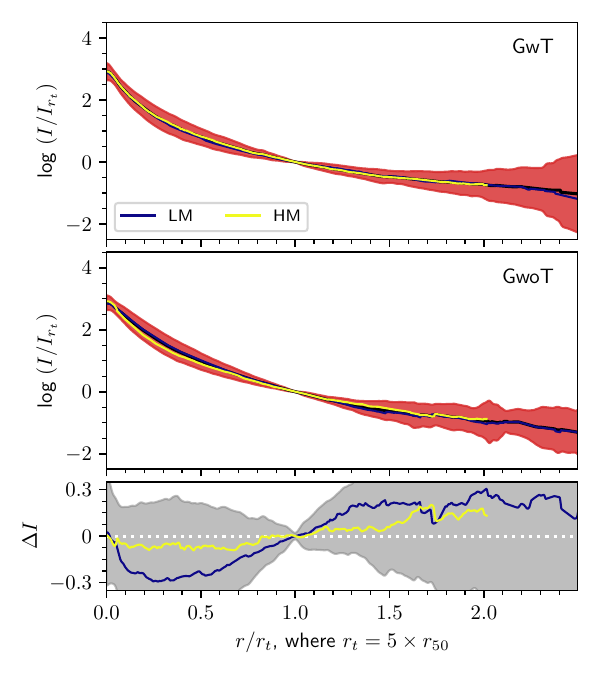}
\caption{Same as Figure~\ref{fig:f8}, but focused only on the impact of stellar mass to examine the cause of the different pattern compared with the observational results. 
\label{fig:f12}}
\end{figure}

As shown in Figure~\ref{fig:f12}, the pronounced bimodal trend is predominantly driven by LM galaxies. We interpret this bimodal response primarily as a geometric artifact arising from merger-induced structural redistribution in dissipationless simulations. As mentioned earlier, the GRT simulation forms less concentrated radial profiles. This effect is particularly severe in GwoT, which lacks the violent relaxation that helps build central concentration in GwT. Furthermore, simulated galaxies in the LM regime experience exaggerated structural inflation: the characteristic scale offset between GwT and GwoT ($\Delta r_{50}/r_{50}$) reaches $\sim$54\% in the simulation, compared to merely $\sim$4\% in the observational sample. Consequently, when the profiles are normalized at $r/r_t=1$, this substantial scale mismatch artificially induces a severe inner trough and exaggerates the outer excess in the LM regime. 

In contrast, in the HM regime, these geometric artifacts are significantly mitigated. Galaxies in the HM regime have deeper potential wells and are thus more resilient to internal structural redistribution and external perturbations, since mergers are predominantly minor accretions, if any. As a result, the inner trough is largely suppressed in the HM regime, revealing a subtle yet distinct outer excess. While this behavior superficially resembles our observational findings in the LD-HM regime, treating it as a direct, quantitative validation of the observations requires strong caution given the restrictive simulation basis. Nonetheless, the emergence of an outer excess in the simulated HM regime suggests that massive systems can indeed retain the structural imprints of accreted material in their outskirts. 

\begin{figure}[t!]
\includegraphics[width=\linewidth]{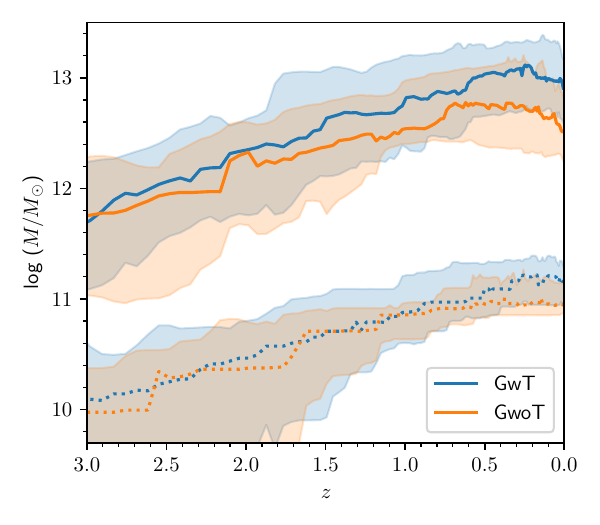}
\caption{Mass growth curves for GwT (blue) and GwoT (orange) as a function of redshift. The solid and dotted lines represent the median dark-matter-halo mass ($M_{200c}$) and stellar mass ($M_\star$), respectively. The shaded regions indicate the 16th--84th percentile range of the data. 
\label{fig:f13}}
\end{figure}

Although complex physical and numerical factors could contribute to the outer positive residuals in simulated galaxies, we directly test whether this structural trend is linked to their underlying mass assembly—the core hypothesis of this study—by tracking the mass growth histories of the simulated galaxies (Figure~\ref{fig:f13}). Across both dark matter and stellar components, GwT systematically experiences accelerated mass growth compared to GwoT throughout their evolutionary histories.\footnote{The decrease in mass near $z=0$ is due to tidal stripping in the densest cluster environments. The dark matter halo is stripped first, while the more deeply bound stellar component remains largely intact.} Importantly, because the dissipationless GRT simulation completely excludes in-situ star formation, these mass growth curves provide an uncontaminated, pristine probe of pure ex-situ accretion. This fundamental divergence in mass growth supports our primary scenario, suggesting that the presence of visible tidal features can indeed reflect a more active and sustained accretion history. 

\subsubsection{Impacts on Detectability of Tidal Debris} \label{sec:disc:valid:impact}
The multiple-projection approach and merger-tree tracking available in simulations enable systematic evaluation of observational biases that affect the detectability of tidal debris. Leveraging these capabilities, we examined several observational and physical factors---projection effects, merger timing and mass ratio, and surface-brightness limit---that affect the recovery of tidal features, thereby providing a quantitative basis for interpreting our observational study. Although external light sources may cause incidental obscuration of tidal debris, we did not account for it because such obscuration is likely to occur only stochastically in very dense regions or near bright stars. 

First, we investigated the projection effect by comparing the three orthogonal views of each parent galaxy. We found that 20 of 70 parent galaxies showed inconsistent tidal-feature identifications across projections, with seven galaxies showing tidal features in only a single projection and 13 visible in two of the three projections. This indicates that a substantial fraction of ``apparent'' GwoT may harbor hidden tidal debris. Therefore, any observational census of tidal features may be a lower bound on the true frequency. It is also worth noting that, although our study focuses on the binary identification of tidal features, characterizing specific morphologies---such as shells, streams, and tails---is likely even more sensitive to these orientation-induced biases. 

\begin{figure}[t!]
\includegraphics[width=\linewidth]{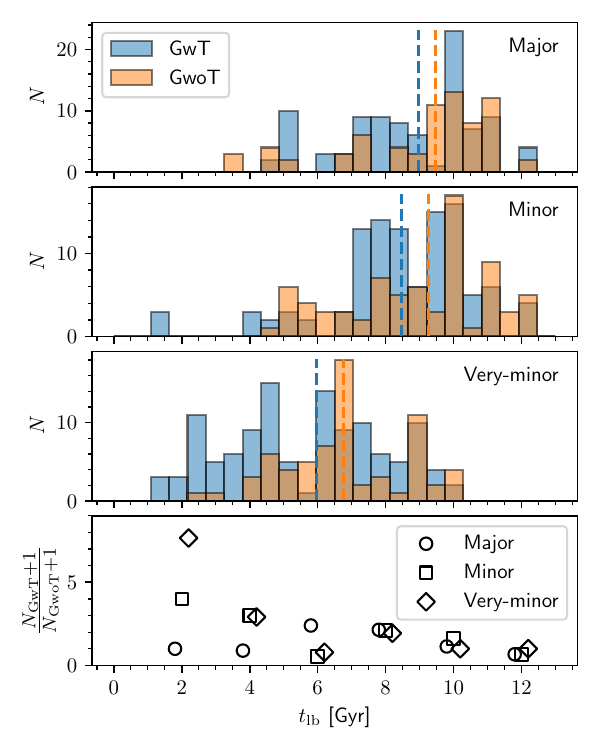}
\caption{Distributions of the lookback time ($t_\mathrm{lb}$) for the most recent merger events in GwT and GwoT. The top three panels show raw event counts for major, minor, and very-minor mergers, respectively, where the vertical dashed lines indicate the median $t_\mathrm{lb}$. According to the Anderson--Darling test, only very-minor mergers are statistically distinct ($p=0.004$). The bottom panel shows the GwT-to-GwoT count ratio, $(N_\mathrm{GwT}+1)/(N_\mathrm{GwoT}+1)$, across coarse $t_\mathrm{lb}$ bins. 
\label{fig:f14}}
\end{figure}

Second, we examined the physical trends underlying the observed tidal features by separately tracking the lookback time ($t_\mathrm{lb}$) of the most recent merger events, categorized as major ($m/M>$~1:3), minor (1:10~$<m/M<$~1:3), or very-minor ($m/M<$~1:10) based on their stellar mass ratios at the moment of interaction. As shown in Figure~\ref{fig:f14}, across all three merger classes, the median $t_\mathrm{lb}$ for GwT is consistently shifted toward more recent epochs by 1--2~Gyr relative to GwoT, reflecting an overall tendency for GwT hosts to experience sustained assembly. Beyond this global offset, two primary trends emerge: (1) both GwT and GwoT have comparable merger histories across all classes in the ancient regime ($t_\mathrm{lb} \gtrsim 6$~Gyr), and (2) the primary divergence appears in the very-minor-merger histories, where GwT shows a clear tendency to sustain accretion more recently ($t_\mathrm{lb} \lesssim 4$~Gyr).\footnote{This timescale is consistent with theoretical estimates for the visibility of tidal debris; for instance, \cite{2019A&A...632A.122M} reported that stellar streams, tails, and shells typically remain detectable for approximately 2--4~Gyr.} 

To more explicitly capture this trend, we calculated the GwT-to-GwoT count ratio---$(N_\mathrm{GwT}+1)/(N_\mathrm{GwoT}+1)$---as a function of lookback time (bottom panel of Figure~\ref{fig:f14}). In the ancient regime, the ratios for all merger classes converge near unity, reflecting the shared early mass assembly of both populations. In contrast, the ratio for minor and very-minor mergers rises at $t_\mathrm{lb} \lesssim 4$~Gyr. Although this recent accretion does not account for the entire GwT---and secondary factors, such as debris survival time and projection effects, may play a role---these findings demonstrate that visible tidal features partially indicate a sustained and active assembly history. 

Finally, we assessed the impact of photometric depth by comparing our mock images with a deeper surface-brightness limit of 30 mag~arcsec$^{-2}$. At this depth, the detection rate of tidal features rises dramatically to $\sim$81\% (171 of 210 realizations). Across all three projected views, only $\sim$11\% (8 of 70 parent galaxies) remain truly devoid of detectable features. This ubiquity of tidal debris indicates that merger remnants are nearly universal in massive ETGs, though their visibility at certain observational limits is selectively governed by the timing of the last merger, the mass of the accreted material, and the dynamical survival time of the debris. That is, our observational GwoT sample is likely contaminated by galaxies harboring faint or obscured debris, and the subtle signals identified in our observational study may represent a conservative lower bound on the true physical response. 

\subsection{Systematics and Caveats} \label{sec:disc:caveats}
While our results provide compelling evidence of a link between detectable tidal features and stellar halo growth, we must address several observational and methodological factors that could affect these findings. 

\paragraph{PSF Effects} The outskirts of surface-brightness profiles are inherently sensitive to observational systematics, particularly the PSF's extended wings. However, we expect PSF-induced contamination to be minor in this study. All profiles were normalized to a consistent physical scale based on $r_{50}$, and we observed no anomalous scatter or upturns at large radii that would indicate the dominance of the PSF's wings. The consistency of the profiles across different masses, environments, and distances further suggests that the detected signal is dominated by the stellar distribution rather than the instrumental response. We also considered the possibility that the PSF core might artificially inflate the measured $r_{50}$. However, given that our sample galaxies have angular sizes significantly larger than the PSF core (typically $\sim$8\arcsec~in radius), the impact of convolution is expected to be negligible; tests with simulated samples indicate a few percent difference at most. 

\paragraph{Sky Subtraction} The background sky model in the DLIS pipeline was computed using the median values within sliding 512-pixel boxes. This method can sometimes overestimate the background level for galaxies with large angular sizes relative to the box size, or for small, high-redshift galaxies in crowded regions \cite[see][]{2017ApJ...834...73Y}, potentially leading to artificial truncation at the galaxy's outskirts during sky subtraction. However, in this work, these effects are largely mitigated because most galaxies in our sample avoid these extreme cases. Indeed, our visual inspection revealed that only a negligible fraction ($\sim$1\%) of the sample showed signs of potential over-subtraction, which is unlikely to affect the main results. Furthermore, because sky oversubtraction would systematically suppress the outer profile, creating a flux deficit rather than an excess, this systematic effect cannot account for the outer excess consistently observed in our GwT sample. 

\paragraph{Imaging Depth} The effective imaging depth of our data is significantly enhanced by azimuthal averaging along isophotes. Because the error for each isophote decreases with the number of contributing pixels, scaling as $N^{-1/2}$, the surface-brightness levels fall below the limit set by the pixel-to-pixel background noise. Indeed, the detection of smooth, continuous outer structures extending well beyond $10\times r_{50}$, without any nonphysical artifacts, confirms that our profiles represent genuine stellar envelopes rather than observational noise or background fluctuations. Additionally, the successful recovery of profiles out to $10\times r_{50}$ indicates that the choice of integration mode in the isophote-fitting procedure did not introduce severe systematic uncertainties, given our primary focus on the overall shapes and slopes of the surface-brightness profiles in the outer regions. 

\paragraph{Identification Completeness and Bias} As discussed in Section~\ref{sec:disc:valid:impact}, the detectability of tidal features is affected by projection effects and photometric depth. Our simulation-based analysis indicates that approximately 30\% of galaxies show inconsistent tidal-feature identification due to projection effects, and nearly 90\% harbor underlying debris at a deeper surface-brightness limit. This suggests that our GwoT sample is likely contaminated by galaxies with hidden, insignificant accretion events. However, because projection effects are largely random and depth-induced incompleteness is likely consistent across the entire sample, these factors are unlikely to introduce systematic biases or false-positive signals. Because such contamination would typically dilute the observed residuals, the subtle signals identified in our observational study may instead represent conservative lower bounds on the true physical response. 

\paragraph{Redshift Evolution} As noted in Section~\ref{sec:method:norm:rmax}, the GwoT sample shows a subtle systematic shift toward higher redshifts than the GwT sample. One might worry that this disparity could introduce an evolutionary bias, with GwT at lower redshifts having more time to accumulate outer structures through late-stage mergers. However, the maximum time difference across the redshift range is only $\sim$0.4~Gyr. Given that the typical dynamical timescales in the outskirts of massive ETGs are likely several Gyr, this interval is physically insufficient to drive significant structural divergence in the stellar halos. 

\paragraph{Robustness of Residuals and Scaling Dependence} Although the residual signals in the outer regions show coherent trends across the mass-density grid, several methodological caveats must be acknowledged. First, the observed differences between GwT and GwoT profiles are confined to specific radial intervals and depend on the radial normalization. Normalizing by characteristic scales such as $r_{50}$ or $r_t$ provides an observationally accessible baseline, but variations in the internal light distribution can subtly affect the alignment of outer profiles. Second, subdividing our sample into a $2\times2$ grid inevitably reduces the number of galaxies per bin, thereby increasing the statistical uncertainty around the median profiles. Given the large galaxy-to-galaxy dispersion in outer surface brightness, our empirical detections should be regarded as suggestive rather than conclusive. Deeper observations and larger, complete samples will be essential to definitively determine whether these localized profile deviations persist beyond observational and sample-size uncertainties. 

\paragraph{Information Loss by Normalization} Normalizing with $r_{50}$ and generating median-stacked profiles can, in principle, smooth out individual structural nuances. This information loss is unavoidable when aggregating a diverse sample. A more discriminative approach for future work would be to perform multi-component decomposition on a galaxy-by-galaxy basis, perhaps anchoring the profiles to each system's intrinsic transition radius. Such an analysis, supported by even deeper imaging data, would enable more accurate discrimination of stellar halos and further refine this study's results. 

\subsection{Implications for Galaxy Evolution} \label{sec:disc:impl}
The structural variations in the outskirts of galaxies identified in this study offer a new perspective on the late-stage assembly of massive ETGs. While these systems are often considered structurally converged \cite[see][]{2018MNRAS.475.3348H, 2020A&A...639A..14S}, with only transient changes induced by certain mergers \cite[see][]{2020ApJ...905..154Y, 2026MNRAS.545f2223G}, our findings suggest that they still exhibit subtle but systematic differentiation driven by their recent assembly histories and cosmic environments. 

A key implication of this study is that the currently observed tidal features—particularly in massive ETGs within lower-density environments—can reflect not merely transient tracers of isolated, stochastic interactions, but may also serve as visible signatures of a more active and sustained mass-assembly history. While not universal across the entire population, our simulation-based analysis supports this scenario by showing that a significant subset of GwT experienced prolonged mass growth, driven by a higher frequency of recent very-minor accretions. These combined findings suggest that visible tidal debris, at least in certain regimes, captures the ongoing manifestation of extended hierarchical assembly. 

However, as emphasized by recent cosmological simulations \cite[e.g.,][]{2024MNRAS.530.4422K}, the detectability of tidal debris remains a complex interplay of intrinsic and extrinsic factors. Although we also recognize that the visibility of tidal debris is a probabilistic outcome---governed by environmental conditions, progenitor mass, merger timing, projection effects, and photometric depth---this suggests that the subtle signals identified in our observational study likely represent only the most prominent manifestations of a much broader underlying phenomenon. Consequently, our results should be considered as a conservative lower bound on the true physical divergence in stellar halo growth. 

Upcoming deep and wide-field surveys, such as the Legacy Survey of Space and Time \cite[][]{2019ApJ...873..111I} and the KASI Deep Rolling Imaging Fast Telescope survey \cite[][]{2026JKAS...59...67K, 2026JKAS...59..117B}, will be essential for achieving the necessary photometric depth ($\gtrsim$30 mag~arcsec$^{-2}$) and for enhancing statistical power. These datasets will allow us to systematically disentangle the effects of detectable tidal debris from the underlying stellar halo growth, thereby enabling definitive validation of the assembly-driven differentiation of massive ETGs and expanding our understanding of how ex-situ processes continue to shape the LSB universe. 

\section{Summary and Conclusion} \label{sec:sum}
In this study, we investigated whether tidal features in massive ETGs are linked to fundamental differences in their extended stellar halo structures. Using a sample of 436 visually classified ETGs from DLIS, we analyzed azimuthally averaged surface-brightness profiles out to $\sim$$10\times r_{50}$. By implementing a multi-dimensional analysis across stellar mass and environmental density, we uncovered subtle but systematic structural variations that are often lost in global averages. 

Our main findings are summarized as follows:
\begin{enumerate}
\item{Identification of Structural Divergence: Contrary to the traditional view of structural convergence, we identified systematic divergence between GwT and GwoT. This differentiation suggests that distinct mass-assembly histories leave a recognizable imprint on their late-stage structural evolution.}
\item{Environment- and Mass-driven Modulation: This structural divergence is most pronounced in the LD-HM and HD-LM regimes. The positive residual in the LD-HM regime suggests that higher-mass systems accelerate ex-situ accretion, whereas lower-density environments serve as a quiet state that preserves the cumulative result of mass assembly. In contrast, the negative residual in the HD-LM regime suggests that lower-mass systems in higher-density environments are likely to undergo tidal stripping in their diffuse outskirts, thereby attenuating ex-situ structural signatures.}
\item{Physical Validation through Simulations: Our results were evaluated with data-matched GRT simulations. The simulated galaxies superficially mirror a part of the observed residual trends. More importantly, tracking their mass growth histories more definitively confirmed that visible tidal features---often triggered by recent very-minor mergers---can be indicative, at least for a subset of galaxies, of an active, sustained mass-assembly history that drives both stellar halo growth and merger-induced structural inflation, rather than isolated, stochastic events.}
\end{enumerate}

We addressed several potential systematics, including PSF wings, projection effects, and photometric depth. Although projection effects and photometric depth can lead to the misidentification of tidal features, our simulation-based analysis indicates that such contamination is largely stochastic and tends to dilute the observed residuals. Consequently, the structural differences identified in this study likely represent conservative lower bounds on the true physical divergence in stellar halo growth. 

In conclusion, tidal features provide a crucial observational link to a galaxy's ongoing mass assembly. Our findings shift the perspective from viewing mergers as transient, disruptive events to recognizing them as visible markers of ongoing structural evolution in the LSB universe. Despite being a depth-limited study, this work provides an integrated analysis of tidal debris and stellar halos, thereby introducing a novel observational pathway for unraveling the complex mass-assembly histories of galaxies. To definitively resolve the full extent of this assembly-driven differentiation, future research must push the boundaries of photometric depth. Upcoming intensive, wide-field surveys will be essential for disentangling the complex interplay among merger history, environmental modulation, and the ultimate growth of the most massive galaxies in the universe. 

\begin{acknowledgments}
We are grateful to an anonymous referee for constructive comments and suggestions.
This research was supported by the Korea Astronomy and Space Science Institute under the R\&D program (Project No. 2026-1-831-03) supervised by the Korea AeroSpace Administration.
This research was supported by Kyungpook National University Research Fund, 2025.
This research was supported by the National Research Foundation of Korea (NRF) grant funded by the Korea government (MSIT) (RS-2025-16064514).
This work made use of Astropy:\footnote{\url{http://www.astropy.org}}, a community-developed core Python package and an ecosystem of tools and resources for astronomy \cite[][]{2013A&A...558A..33A, 2018AJ....156..123A, 2022ApJ...935..167A}.
This research made use of Photutils, an Astropy package for detection and photometry of astronomical sources \cite[][]{larry_bradley_2022_6825092}.
This research is also supported by Uncertainties\footnote{\url{http://pythonhosted.org/uncertainties}}: a Python package for calculations with uncertainties, by Eric O. LEBIGOT.

\end{acknowledgments}

\begin{contribution}

WB conceived this study, led the data analysis, and took primary responsibility for writing the manuscript.
YY provided the galaxy catalog and the motivation for this work based on their prior study, supported data collection, and contributed to revising the manuscript.
KC provided the simulation data, assisted with the theoretical interpretation, and helped revise the manuscript.
JK secured funding for this project and contributed to revising the manuscript.
JWK contributed to scientific discussions and to revising the manuscript.



\end{contribution}

%

\software{Astropy \cite[][]{2013A&A...558A..33A, 2018AJ....156..123A, 2022ApJ...935..167A},
Numpy \cite[][]{2020Natur.585..357H},
Photutils \cite[][]{larry_bradley_2022_6825092}
Scipy \cite[][]{2020NatMe..17..261V},
SExtractor \cite[][]{1996A&AS..117..393B}
}




\bibliography{ms}{}

@article{2003MNRAS.344.1000B,
	adsurl = {https://ui.adsabs.harvard.edu/abs/2003MNRAS.344.1000B},
	archiveprefix = {arXiv},
	author = {{Bruzual}, G. and {Charlot}, S.},
	doi = {10.1046/j.1365-8711.2003.06897.x},
	eprint = {astro-ph/0309134},
	journal = {\mnras},
	month = oct,
	number = {4},
	pages = {1000-1028},
	primaryclass = {astro-ph},
	title = {{Stellar population synthesis at the resolution of 2003}},
	volume = {344},
	year = 2003}

@article{2016MNRAS.456.2140M,
	adsurl = {https://ui.adsabs.harvard.edu/abs/2016MNRAS.456.2140M},
	archiveprefix = {arXiv},
	author = {{Ma}, Xiangcheng and {Hopkins}, Philip F. and {Faucher-Gigu{\`e}re}, Claude-Andr{\'e} and {Zolman}, Nick and {Muratov}, Alexander L. and {Kere{\v{s}}}, Du{\v{s}}an and {Quataert}, Eliot},
	doi = {10.1093/mnras/stv2659},
	eprint = {1504.02097},
	journal = {\mnras},
	month = feb,
	number = {2},
	pages = {2140-2156},
	primaryclass = {astro-ph.GA},
	title = {{The origin and evolution of the galaxy mass-metallicity relation}},
	volume = {456},
	year = 2016}

@article{2026ApJ...996..145C,
	adsurl = {https://ui.adsabs.harvard.edu/abs/2026ApJ...996..145C},
	archiveprefix = {arXiv},
	author = {{Chun}, Kyungwon and {Shin}, Jihye and {Ko}, Jongwan and {Smith}, Rory and {Park}, So-Myoung and {Nam}, Songhee},
	doi = {10.3847/1538-4357/ae0cc7},
	eid = {145},
	eprint = {2509.22802},
	journal = {\apj},
	month = jan,
	number = {2},
	pages = {145},
	primaryclass = {astro-ph.GA},
	title = {{The Role of Preprocessing in Tidal Feature Formation within Galaxy Clusters}},
	volume = {996},
	year = 2026}

@article{2007ApJ...658..710N,
	adsurl = {https://ui.adsabs.harvard.edu/abs/2007ApJ...658..710N},
	archiveprefix = {arXiv},
	author = {{Naab}, Thorsten and {Johansson}, Peter H. and {Ostriker}, Jeremiah P. and {Efstathiou}, George},
	doi = {10.1086/510841},
	eprint = {astro-ph/0512235},
	journal = {\apj},
	month = apr,
	number = {2},
	pages = {710-720},
	primaryclass = {astro-ph},
	title = {{Formation of Early-Type Galaxies from Cosmological Initial Conditions}},
	volume = {658},
	year = 2007}

@article{2009ApJS..182..216K,
	adsurl = {https://ui.adsabs.harvard.edu/abs/2009ApJS..182..216K},
	archiveprefix = {arXiv},
	author = {{Kormendy}, John and {Fisher}, David B. and {Cornell}, Mark E. and {Bender}, Ralf},
	doi = {10.1088/0067-0049/182/1/216},
	eprint = {0810.1681},
	journal = {\apjs},
	month = may,
	number = {1},
	pages = {216-309},
	primaryclass = {astro-ph},
	title = {{Structure and Formation of Elliptical and Spheroidal Galaxies}},
	volume = {182},
	year = 2009}

@article{2007ApJ...658...65C,
	adsurl = {https://ui.adsabs.harvard.edu/abs/2007ApJ...658...65C},
	archiveprefix = {arXiv},
	author = {{Ciotti}, L. and {Lanzoni}, B. and {Volonteri}, M.},
	doi = {10.1086/510773},
	eprint = {astro-ph/0611328},
	journal = {\apj},
	month = mar,
	number = {1},
	pages = {65-77},
	primaryclass = {astro-ph},
	title = {{The Importance of Dry and Wet Merging on the Formation and Evolution of Elliptical Galaxies}},
	volume = {658},
	year = 2007}

@article{2017ApJ...834...73Y,
	adsurl = {https://ui.adsabs.harvard.edu/abs/2017ApJ...834...73Y},
	archiveprefix = {arXiv},
	author = {{Yoon}, Yongmin and {Im}, Myungshin and {Kim}, Jae-Woo},
	doi = {10.3847/1538-4357/834/1/73},
	eid = {73},
	eprint = {1612.07945},
	journal = {\apj},
	month = jan,
	number = {1},
	pages = {73},
	primaryclass = {astro-ph.GA},
	title = {{Massive Galaxies Are Larger in Dense Environments: Environmental Dependence of Mass-Size Relation of Early-type Galaxies}},
	volume = {834},
	year = 2017}

@article{2019A&A...632A.122M,
	adsurl = {https://ui.adsabs.harvard.edu/abs/2019A&A...632A.122M},
	archiveprefix = {arXiv},
	author = {{Mancillas}, Brisa and {Duc}, Pierre-Alain and {Combes}, Fran{\c{c}}oise and {Bournaud}, Fr{\'e}d{\'e}ric and {Emsellem}, Eric and {Martig}, Marie and {Michel-Dansac}, Leo},
	doi = {10.1051/0004-6361/201936320},
	eid = {A122},
	eprint = {1909.07500},
	journal = {\aap},
	month = dec,
	pages = {A122},
	primaryclass = {astro-ph.GA},
	title = {{Probing the merger history of red early-type galaxies with their faint stellar substructures}},
	volume = {632},
	year = 2019}

@article{2026JKAS...59..117B,
	adsurl = {https://ui.adsabs.harvard.edu/abs/2026JKAS...59..117B},
	archiveprefix = {arXiv},
	author = {{Byun}, Woowon and {Yoon}, Yongmin and {Ko}, Jongwan and {Lee}, Yun Hee and {Lee}, Gain and {Hwang}, Ho Seong and {Sabiu}, Cristiano G. and {Seon}, Kwang-il and {Chun}, Kyungwon and {Shin}, Jihye and {Rhee}, Jinsu and {Kim}, Jae-Woo and {Yoo}, Jaewon and {Lee}, Jaehyun and {Chun}, Sang-Hyun and {Park}, Hong Soo and {Yang}, Soung-Chul and {Hong}, Sungryong and {Shin}, Jeehye and {Kim}, Hyowon},
	doi = {10.5303/JKAS.2026.59.1.117},
	eprint = {2602.08283},
	journal = {Journal of Korean Astronomical Society},
	month = mar,
	pages = {117-139},
	primaryclass = {astro-ph.GA},
	title = {{K-DRIFT Science Theme: Galaxies in the Faint Universe}},
	volume = {59},
	year = 2026}

@article{2026JKAS...59...67K,
	adsurl = {https://ui.adsabs.harvard.edu/abs/2026JKAS...59...67K},
	archiveprefix = {arXiv},
	author = {{Ko}, Jongwan and {Byun}, Woowon and {Seon}, Kwang-Il and {Kim}, Jihun and {Kim}, Yunjong and {Kim}, Daewook and {Chang}, Seunghyuk and {Kim}, Dohoon and {Kweon Moon}, Il and {Kwon}, Hyuksun and {Kim}, Yeonsik and {Ahn}, Kyohoon and {Lee}, Gayoung and {Lee}, Yongseok and {Lee}, Sangmin and {Cha}, Sang-Mok and {Kim}, Dong-Jin and {Park}, Kyusu and {Yoo}, Jaewon and {Kim}, Jae-Woo and {Shin}, Jihye and {Chun}, Sang-Hyun and {Yoon}, Yongmin and {Lee}, Jaehyun and {Chun}, Kyungwon and {Rhee}, Jinsu and {Hong}, Sungryong and {Park}, Jongyeob and {Jeon}, Young-Beom and {Sung}, Eon-Chang and {Park}, Hong Soo and {Kim}, Seonwoo Kim and {Bahk}, GyeongGon and {Yeon}, Seri},
	doi = {10.5303/JKAS.2026.59.1.67},
	eprint = {2510.22250},
	journal = {Journal of Korean Astronomical Society},
	month = mar,
	pages = {67-80},
	primaryclass = {astro-ph.GA},
	title = {{K-DRIFT: Unveiling New Imagery of the Hidden Universe}},
	volume = {59},
	year = 2026}

@article{2025A&A...699A..93T,
	adsurl = {https://ui.adsabs.harvard.edu/abs/2025A&A...699A..93T},
	archiveprefix = {arXiv},
	author = {{Tau}, Elisa A. and {Monachesi}, Antonela and {Gomez}, Facundo A. and {Grand}, Robert J.~J. and {Pakmor}, R{\"u}diger and {van de Voort}, Freeke and {Gonzalez-Jara}, Jenny and {Tissera}, Patricia B. and {Marinacci}, Federico and {Bieri}, Rebekka},
	doi = {10.1051/0004-6361/202453488},
	eid = {A93},
	eprint = {2412.13807},
	journal = {\aap},
	month = jul,
	pages = {A93},
	primaryclass = {astro-ph.GA},
	title = {{The role of accreted and in situ populations in shaping the stellar halos of low-mass galaxies}},
	volume = {699},
	year = 2025}

@article{2026MNRAS.545f2223G,
	adsurl = {https://ui.adsabs.harvard.edu/abs/2026MNRAS.545f2223G},
	archiveprefix = {arXiv},
	author = {{Gordon}, Alexander J. and {Ferguson}, Annette M.~N. and {Mann}, Robert G. and {Wild}, Vivienne},
	doi = {10.1093/mnras/staf2223},
	eid = {staf2223},
	eprint = {2507.21050},
	journal = {\mnras},
	month = feb,
	number = {4},
	pages = {staf2223},
	primaryclass = {astro-ph.GA},
	title = {{Linking enhanced star formation and quenching to faint tidal features in galaxies}},
	volume = {545},
	year = 2026}

@article{2011ApJ...739...20C,
	adsurl = {https://ui.adsabs.harvard.edu/abs/2011ApJ...739...20C},
	archiveprefix = {arXiv},
	author = {{Courteau}, St{\'e}phane and {Widrow}, Lawrence M. and {McDonald}, Michael and {Guhathakurta}, Puragra and {Gilbert}, Karoline M. and {Zhu}, Yucong and {Beaton}, Rachael Lynn and {Majewski}, Steven R.},
	doi = {10.1088/0004-637X/739/1/20},
	eid = {20},
	eprint = {1106.3564},
	journal = {\apj},
	month = sep,
	number = {1},
	pages = {20},
	primaryclass = {astro-ph.CO},
	title = {{The Luminosity Profile and Structural Parameters of the Andromeda Galaxy}},
	volume = {739},
	year = 2011}

@article{2012arXiv1204.3082B,
	adsurl = {https://ui.adsabs.harvard.edu/abs/2012arXiv1204.3082B},
	archiveprefix = {arXiv},
	author = {{Bakos}, Judit and {Trujillo}, Ignacio},
	doi = {10.48550/arXiv.1204.3082},
	eid = {arXiv:1204.3082},
	eprint = {1204.3082},
	journal = {arXiv e-prints},
	month = apr,
	pages = {arXiv:1204.3082},
	primaryclass = {astro-ph.CO},
	title = {{Deep Surface Brightness Profiles of Spiral Galaxies from SDSS Stripe82: Touching Stellar Halos}},
	year = 2012}

@article{2020Natur.585..357H,
	adsurl = {https://ui.adsabs.harvard.edu/abs/2020Natur.585..357H},
	archiveprefix = {arXiv},
	author = {{Harris}, Charles R. and {Millman}, K. Jarrod and {van der Walt}, St{\'e}fan J. and {Gommers}, Ralf and {Virtanen}, Pauli and {Cournapeau}, David and {Wieser}, Eric and {Taylor}, Julian and {Berg}, Sebastian and {Smith}, Nathaniel J. and {Kern}, Robert and {Picus}, Matti and {Hoyer}, Stephan and {van Kerkwijk}, Marten H. and {Brett}, Matthew and {Haldane}, Allan and {del R{\'\i}o}, Jaime Fern{\'a}ndez and {Wiebe}, Mark and {Peterson}, Pearu and {G{\'e}rard-Marchant}, Pierre and {Sheppard}, Kevin and {Reddy}, Tyler and {Weckesser}, Warren and {Abbasi}, Hameer and {Gohlke}, Christoph and {Oliphant}, Travis E.},
	doi = {10.1038/s41586-020-2649-2},
	eprint = {2006.10256},
	journal = {\nat},
	month = sep,
	number = {7825},
	pages = {357-362},
	primaryclass = {cs.MS},
	title = {{Array programming with NumPy}},
	volume = {585},
	year = 2020}

@article{2019ApJ...873..111I,
	adsurl = {https://ui.adsabs.harvard.edu/abs/2019ApJ...873..111I},
	archiveprefix = {arXiv},
	author = {{Ivezi{\'c}}, {\v{Z}}eljko and {Kahn}, Steven M. and {Tyson}, J. Anthony and {Abel}, Bob and {Acosta}, Emily and {Allsman}, Robyn and {Alonso}, David and {AlSayyad}, Yusra and {Anderson}, Scott F. and {Andrew}, John and {Angel}, James Roger P. and {Angeli}, George Z. and {Ansari}, Reza and {Antilogus}, Pierre and {Araujo}, Constanza and {Armstrong}, Robert and {Arndt}, Kirk T. and {Astier}, Pierre and {Aubourg}, {\'E}ric and {Auza}, Nicole and {Axelrod}, Tim S. and {Bard}, Deborah J. and {Barr}, Jeff D. and {Barrau}, Aurelian and {Bartlett}, James G. and {Bauer}, Amanda E. and {Bauman}, Brian J. and {Baumont}, Sylvain and {Bechtol}, Ellen and {Bechtol}, Keith and {Becker}, Andrew C. and {Becla}, Jacek and {Beldica}, Cristina and {Bellavia}, Steve and {Bianco}, Federica B. and {Biswas}, Rahul and {Blanc}, Guillaume and {Blazek}, Jonathan and {Blandford}, Roger D. and {Bloom}, Josh S. and {Bogart}, Joanne and {Bond}, Tim W. and {Booth}, Michael T. and {Borgland}, Anders W. and {Borne}, Kirk and {Bosch}, James F. and {Boutigny}, Dominique and {Brackett}, Craig A. and {Bradshaw}, Andrew and {Brandt}, William Nielsen and {Brown}, Michael E. and {Bullock}, James S. and {Burchat}, Patricia and {Burke}, David L. and {Cagnoli}, Gianpietro and {Calabrese}, Daniel and {Callahan}, Shawn and {Callen}, Alice L. and {Carlin}, Jeffrey L. and {Carlson}, Erin L. and {Chandrasekharan}, Srinivasan and {Charles-Emerson}, Glenaver and {Chesley}, Steve and {Cheu}, Elliott C. and {Chiang}, Hsin-Fang and {Chiang}, James and {Chirino}, Carol and {Chow}, Derek and {Ciardi}, David R. and {Claver}, Charles F. and {Cohen-Tanugi}, Johann and {Cockrum}, Joseph J. and {Coles}, Rebecca and {Connolly}, Andrew J. and {Cook}, Kem H. and {Cooray}, Asantha and {Covey}, Kevin R. and {Cribbs}, Chris and {Cui}, Wei and {Cutri}, Roc and {Daly}, Philip N. and {Daniel}, Scott F. and {Daruich}, Felipe and {Daubard}, Guillaume and {Daues}, Greg and {Dawson}, William and {Delgado}, Francisco and {Dellapenna}, Alfred and {de Peyster}, Robert and {de Val-Borro}, Miguel and {Digel}, Seth W. and {Doherty}, Peter and {Dubois}, Richard and {Dubois-Felsmann}, Gregory P. and {Durech}, Josef and {Economou}, Frossie and {Eifler}, Tim and {Eracleous}, Michael and {Emmons}, Benjamin L. and {Fausti Neto}, Angelo and {Ferguson}, Henry and {Figueroa}, Enrique and {Fisher-Levine}, Merlin and {Focke}, Warren and {Foss}, Michael D. and {Frank}, James and {Freemon}, Michael D. and {Gangler}, Emmanuel and {Gawiser}, Eric and {Geary}, John C. and {Gee}, Perry and {Geha}, Marla and {Gessner}, Charles J.~B. and {Gibson}, Robert R. and {Gilmore}, D. Kirk and {Glanzman}, Thomas and {Glick}, William and {Goldina}, Tatiana and {Goldstein}, Daniel A. and {Goodenow}, Iain and {Graham}, Melissa L. and {Gressler}, William J. and {Gris}, Philippe and {Guy}, Leanne P. and {Guyonnet}, Augustin and {Haller}, Gunther and {Harris}, Ron and {Hascall}, Patrick A. and {Haupt}, Justine and {Hernandez}, Fabio and {Herrmann}, Sven and {Hileman}, Edward and {Hoblitt}, Joshua and {Hodgson}, John A. and {Hogan}, Craig and {Howard}, James D. and {Huang}, Dajun and {Huffer}, Michael E. and {Ingraham}, Patrick and {Innes}, Walter R. and {Jacoby}, Suzanne H. and {Jain}, Bhuvnesh and {Jammes}, Fabrice and {Jee}, M. James and {Jenness}, Tim and {Jernigan}, Garrett and {Jevremovi{\'c}}, Darko and {Johns}, Kenneth and {Johnson}, Anthony S. and {Johnson}, Margaret W.~G. and {Jones}, R. Lynne and {Juramy-Gilles}, Claire and {Juri{\'c}}, Mario and {Kalirai}, Jason S. and {Kallivayalil}, Nitya J. and {Kalmbach}, Bryce and {Kantor}, Jeffrey P. and {Karst}, Pierre and {Kasliwal}, Mansi M. and {Kelly}, Heather and {Kessler}, Richard and {Kinnison}, Veronica and {Kirkby}, David and {Knox}, Lloyd and {Kotov}, Ivan V. and {Krabbendam}, Victor L. and {Krughoff}, K. Simon and {Kub{\'a}nek}, Petr and {Kuczewski}, John and {Kulkarni}, Shri and {Ku}, John and {Kurita}, Nadine R. and {Lage}, Craig S. and {Lambert}, Ron and {Lange}, Travis and {Langton}, J. Brian and {Le Guillou}, Laurent and {Levine}, Deborah and {Liang}, Ming and {Lim}, Kian-Tat and {Lintott}, Chris J. and {Long}, Kevin E. and {Lopez}, Margaux and {Lotz}, Paul J. and {Lupton}, Robert H. and {Lust}, Nate B. and {MacArthur}, Lauren A. and {Mahabal}, Ashish and {Mandelbaum}, Rachel and {Markiewicz}, Thomas W. and {Marsh}, Darren S. and {Marshall}, Philip J. and {Marshall}, Stuart and {May}, Morgan and {McKercher}, Robert and {McQueen}, Michelle and {Meyers}, Joshua and {Migliore}, Myriam and {Miller}, Michelle and {Mills}, David J.},
	doi = {10.3847/1538-4357/ab042c},
	eid = {111},
	eprint = {0805.2366},
	journal = {\apj},
	month = mar,
	number = {2},
	pages = {111},
	primaryclass = {astro-ph},
	title = {{LSST: From Science Drivers to Reference Design and Anticipated Data Products}},
	volume = {873},
	year = 2019}

@article{2020A&A...639A..14S,
	adsurl = {https://ui.adsabs.harvard.edu/abs/2020A&A...639A..14S},
	archiveprefix = {arXiv},
	author = {{Spavone}, M. and {Iodice}, E. and {van de Ven}, G. and {Falc{\'o}n-Barroso}, J. and {Raj}, M.~A. and {Hilker}, M. and {Peletier}, R.~P. and {Capaccioli}, M. and {Mieske}, S. and {Venhola}, A. and {Napolitano}, N.~R. and {Cantiello}, M. and {Paolillo}, M. and {Schipani}, P.},
	doi = {10.1051/0004-6361/202038015},
	eid = {A14},
	eprint = {2005.00025},
	journal = {\aap},
	month = jul,
	pages = {A14},
	primaryclass = {astro-ph.GA},
	title = {{The Fornax Deep Survey with VST. VIII. Connecting the accretion history with the cluster density}},
	volume = {639},
	year = 2020}

@article{2018MNRAS.475.3348H,
	adsurl = {https://ui.adsabs.harvard.edu/abs/2018MNRAS.475.3348H},
	archiveprefix = {arXiv},
	author = {{Huang}, Song and {Leauthaud}, Alexie and {Greene}, Jenny E. and {Bundy}, Kevin and {Lin}, Yen-Ting and {Tanaka}, Masayuki and {Miyazaki}, Satoshi and {Komiyama}, Yutaka},
	doi = {10.1093/mnras/stx3200},
	eprint = {1707.01904},
	journal = {\mnras},
	month = apr,
	number = {3},
	pages = {3348-3368},
	primaryclass = {astro-ph.GA},
	title = {{Individual stellar haloes of massive galaxies measured to 100 kpc at 0.3 < z < 0.5 using Hyper Suprime-Cam}},
	volume = {475},
	year = 2018}

@article{2000AJ....120.1579Y,
	adsurl = {https://ui.adsabs.harvard.edu/abs/2000AJ....120.1579Y},
	archiveprefix = {arXiv},
	author = {{York}, Donald G. and {Adelman}, J. and {Anderson}, Jr., John E. and {Anderson}, Scott F. and {Annis}, James and {Bahcall}, Neta A. and {Bakken}, J.~A. and {Barkhouser}, Robert and {Bastian}, Steven and {Berman}, Eileen and {Boroski}, William N. and {Bracker}, Steve and {Briegel}, Charlie and {Briggs}, John W. and {Brinkmann}, J. and {Brunner}, Robert and {Burles}, Scott and {Carey}, Larry and {Carr}, Michael A. and {Castander}, Francisco J. and {Chen}, Bing and {Colestock}, Patrick L. and {Connolly}, A.~J. and {Crocker}, J.~H. and {Csabai}, Istv{\'a}n and {Czarapata}, Paul C. and {Davis}, John Eric and {Doi}, Mamoru and {Dombeck}, Tom and {Eisenstein}, Daniel and {Ellman}, Nancy and {Elms}, Brian R. and {Evans}, Michael L. and {Fan}, Xiaohui and {Federwitz}, Glenn R. and {Fiscelli}, Larry and {Friedman}, Scott and {Frieman}, Joshua A. and {Fukugita}, Masataka and {Gillespie}, Bruce and {Gunn}, James E. and {Gurbani}, Vijay K. and {de Haas}, Ernst and {Haldeman}, Merle and {Harris}, Frederick H. and {Hayes}, J. and {Heckman}, Timothy M. and {Hennessy}, G.~S. and {Hindsley}, Robert B. and {Holm}, Scott and {Holmgren}, Donald J. and {Huang}, Chi-hao and {Hull}, Charles and {Husby}, Don and {Ichikawa}, Shin-Ichi and {Ichikawa}, Takashi and {Ivezi{\'c}}, {\v{Z}}eljko and {Kent}, Stephen and {Kim}, Rita S.~J. and {Kinney}, E. and {Klaene}, Mark and {Kleinman}, A.~N. and {Kleinman}, S. and {Knapp}, G.~R. and {Korienek}, John and {Kron}, Richard G. and {Kunszt}, Peter Z. and {Lamb}, D.~Q. and {Lee}, B. and {Leger}, R. French and {Limmongkol}, Siriluk and {Lindenmeyer}, Carl and {Long}, Daniel C. and {Loomis}, Craig and {Loveday}, Jon and {Lucinio}, Rich and {Lupton}, Robert H. and {MacKinnon}, Bryan and {Mannery}, Edward J. and {Mantsch}, P.~M. and {Margon}, Bruce and {McGehee}, Peregrine and {McKay}, Timothy A. and {Meiksin}, Avery and {Merelli}, Aronne and {Monet}, David G. and {Munn}, Jeffrey A. and {Narayanan}, Vijay K. and {Nash}, Thomas and {Neilsen}, Eric and {Neswold}, Rich and {Newberg}, Heidi Jo and {Nichol}, R.~C. and {Nicinski}, Tom and {Nonino}, Mario and {Okada}, Norio and {Okamura}, Sadanori and {Ostriker}, Jeremiah P. and {Owen}, Russell and {Pauls}, A. George and {Peoples}, John and {Peterson}, R.~L. and {Petravick}, Donald and {Pier}, Jeffrey R. and {Pope}, Adrian and {Pordes}, Ruth and {Prosapio}, Angela and {Rechenmacher}, Ron and {Quinn}, Thomas R. and {Richards}, Gordon T. and {Richmond}, Michael W. and {Rivetta}, Claudio H. and {Rockosi}, Constance M. and {Ruthmansdorfer}, Kurt and {Sandford}, Dale and {Schlegel}, David J. and {Schneider}, Donald P. and {Sekiguchi}, Maki and {Sergey}, Gary and {Shimasaku}, Kazuhiro and {Siegmund}, Walter A. and {Smee}, Stephen and {Smith}, J. Allyn and {Snedden}, S. and {Stone}, R. and {Stoughton}, Chris and {Strauss}, Michael A. and {Stubbs}, Christopher and {SubbaRao}, Mark and {Szalay}, Alexander S. and {Szapudi}, Istvan and {Szokoly}, Gyula P. and {Thakar}, Anirudda R. and {Tremonti}, Christy and {Tucker}, Douglas L. and {Uomoto}, Alan and {Vanden Berk}, Dan and {Vogeley}, Michael S. and {Waddell}, Patrick and {Wang}, Shu-i. and {Watanabe}, Masaru and {Weinberg}, David H. and {Yanny}, Brian and {Yasuda}, Naoki and {SDSS Collaboration}},
	doi = {10.1086/301513},
	eprint = {astro-ph/0006396},
	journal = {\aj},
	month = sep,
	number = {3},
	pages = {1579-1587},
	primaryclass = {astro-ph},
	title = {{The Sloan Digital Sky Survey: Technical Summary}},
	volume = {120},
	year = 2000}

@article{2005ApJ...635..931B,
	adsurl = {https://ui.adsabs.harvard.edu/abs/2005ApJ...635..931B},
	archiveprefix = {arXiv},
	author = {{Bullock}, James S. and {Johnston}, Kathryn V.},
	doi = {10.1086/497422},
	eprint = {astro-ph/0506467},
	journal = {\apj},
	month = dec,
	number = {2},
	pages = {931-949},
	primaryclass = {astro-ph},
	title = {{Tracing Galaxy Formation with Stellar Halos. I. Methods}},
	volume = {635},
	year = 2005}

@inproceedings{2004IAUS..217..390M,
	adsurl = {https://ui.adsabs.harvard.edu/abs/2004IAUS..217..390M},
	archiveprefix = {arXiv},
	author = {{Mihos}, J.~C.},
	booktitle = {Recycling Intergalactic and Interstellar Matter},
	doi = {10.48550/arXiv.astro-ph/0401557},
	editor = {{Duc}, Pierre-Alain and {Braine}, Jonathan and {Brinks}, Elias},
	eprint = {astro-ph/0401557},
	month = jun,
	pages = {390},
	primaryclass = {astro-ph},
	series = {IAU Symposium},
	title = {{The Evolution of Tidal Debris}},
	volume = {217},
	year = 2004}

@article{2012A&A...538A..21S,
	adsurl = {https://ui.adsabs.harvard.edu/abs/2012A&A...538A..21S},
	archiveprefix = {arXiv},
	author = {{Schuster}, W.~J. and {Moreno}, E. and {Nissen}, P.~E. and {Pichardo}, B.},
	doi = {10.1051/0004-6361/201118035},
	eid = {A21},
	eprint = {1111.4026},
	journal = {\aap},
	month = feb,
	pages = {A21},
	primaryclass = {astro-ph.GA},
	title = {{Two distinct halo populations in the solar neighborhood. III. Evidence from stellar ages and orbital parameters}},
	volume = {538},
	year = 2012}

@article{2007Natur.450.1020C,
	adsurl = {https://ui.adsabs.harvard.edu/abs/2007Natur.450.1020C},
	archiveprefix = {arXiv},
	author = {{Carollo}, Daniela and {Beers}, Timothy C. and {Lee}, Young Sun and {Chiba}, Masashi and {Norris}, John E. and {Wilhelm}, Ronald and {Sivarani}, Thirupathi and {Marsteller}, Brian and {Munn}, Jeffrey A. and {Bailer-Jones}, Coryn A.~L. and {Fiorentin}, Paola Re and {York}, Donald G.},
	doi = {10.1038/nature06460},
	eprint = {0706.3005},
	journal = {\nat},
	month = dec,
	number = {7172},
	pages = {1020-1025},
	primaryclass = {astro-ph},
	title = {{Two stellar components in the halo of the Milky Way}},
	volume = {450},
	year = 2007}

@article{2010MNRAS.406..744C,
	adsurl = {https://ui.adsabs.harvard.edu/abs/2010MNRAS.406..744C},
	archiveprefix = {arXiv},
	author = {{Cooper}, A.~P. and {Cole}, S. and {Frenk}, C.~S. and {White}, S.~D.~M. and {Helly}, J. and {Benson}, A.~J. and {De Lucia}, G. and {Helmi}, A. and {Jenkins}, A. and {Navarro}, J.~F. and {Springel}, V. and {Wang}, J.},
	doi = {10.1111/j.1365-2966.2010.16740.x},
	eprint = {0910.3211},
	journal = {\mnras},
	month = aug,
	number = {2},
	pages = {744-766},
	primaryclass = {astro-ph.GA},
	title = {{Galactic stellar haloes in the CDM model}},
	volume = {406},
	year = 2010}

@article{2013ApJ...763..113D,
	adsurl = {https://ui.adsabs.harvard.edu/abs/2013ApJ...763..113D},
	archiveprefix = {arXiv},
	author = {{Deason}, A.~J. and {Belokurov}, V. and {Evans}, N.~W. and {Johnston}, K.~V.},
	doi = {10.1088/0004-637X/763/2/113},
	eid = {113},
	eprint = {1210.4929},
	journal = {\apj},
	month = feb,
	number = {2},
	pages = {113},
	primaryclass = {astro-ph.GA},
	title = {{Broken and Unbroken: The Milky Way and M31 Stellar Halos}},
	volume = {763},
	year = 2013}

@article{2013MNRAS.433.2986W,
	adsurl = {https://ui.adsabs.harvard.edu/abs/2013MNRAS.433.2986W},
	archiveprefix = {arXiv},
	author = {{Wilman}, David J. and {Fontanot}, Fabio and {De Lucia}, Gabriella and {Erwin}, Peter and {Monaco}, Pierluigi},
	doi = {10.1093/mnras/stt941},
	eprint = {1305.7163},
	journal = {\mnras},
	month = aug,
	number = {4},
	pages = {2986-3004},
	primaryclass = {astro-ph.CO},
	title = {{The hierarchical origins of observed galaxy morphology}},
	volume = {433},
	year = 2013}

@article{2006MNRAS.366..499D,
	adsurl = {https://ui.adsabs.harvard.edu/abs/2006MNRAS.366..499D},
	archiveprefix = {arXiv},
	author = {{De Lucia}, Gabriella and {Springel}, Volker and {White}, Simon D.~M. and {Croton}, Darren and {Kauffmann}, Guinevere},
	doi = {10.1111/j.1365-2966.2005.09879.x},
	eprint = {astro-ph/0509725},
	journal = {\mnras},
	month = feb,
	number = {2},
	pages = {499-509},
	primaryclass = {astro-ph},
	title = {{The formation history of elliptical galaxies}},
	volume = {366},
	year = 2006}

@article{2004ApJ...616..192C,
	adsurl = {https://ui.adsabs.harvard.edu/abs/2004ApJ...616..192C},
	archiveprefix = {arXiv},
	author = {{Christlein}, Daniel and {Zabludoff}, Ann I.},
	doi = {10.1086/424909},
	eprint = {astro-ph/0408036},
	journal = {\apj},
	month = nov,
	number = {1},
	pages = {192-198},
	primaryclass = {astro-ph},
	title = {{Can Early-Type Galaxies Evolve from the Fading of the Disks of Late-Type Galaxies?}},
	volume = {616},
	year = 2004}

@article{1996MNRAS.283.1361B,
	adsurl = {https://ui.adsabs.harvard.edu/abs/1996MNRAS.283.1361B},
	archiveprefix = {arXiv},
	author = {{Baugh}, C.~M. and {Cole}, S. and {Frenk}, C.~S.},
	doi = {10.1093/mnras/283.4.1361},
	eprint = {astro-ph/9602085},
	journal = {\mnras},
	month = dec,
	number = {4},
	pages = {1361-1378},
	primaryclass = {astro-ph},
	title = {{Evolution of the Hubble sequence in hierarchical models for galaxy formation}},
	volume = {283},
	year = 1996}

@article{2016A&A...588A..79L,
	adsurl = {https://ui.adsabs.harvard.edu/abs/2016A&A...588A..79L},
	archiveprefix = {arXiv},
	author = {{Lacerna}, I. and {Hern{\'a}ndez-Toledo}, H.~M. and {Avila-Reese}, V. and {Abonza-Sane}, J. and {del Olmo}, A.},
	doi = {10.1051/0004-6361/201527844},
	eid = {A79},
	eprint = {1511.08809},
	journal = {\aap},
	month = apr,
	pages = {A79},
	primaryclass = {astro-ph.GA},
	title = {{Isolated elliptical galaxies in the local Universe}},
	volume = {588},
	year = 2016}

@article{2007MNRAS.377.1772R,
	adsurl = {https://ui.adsabs.harvard.edu/abs/2007MNRAS.377.1772R},
	archiveprefix = {arXiv},
	author = {{Reda}, Fatma M. and {Proctor}, Robert N. and {Forbes}, Duncan A. and {Hau}, George K.~T. and {Larsen}, S{\o}ren S.},
	doi = {10.1111/j.1365-2966.2007.11755.x},
	eprint = {astro-ph/0703545},
	journal = {\mnras},
	month = jun,
	number = {4},
	pages = {1772-1784},
	primaryclass = {astro-ph},
	title = {{The spatially resolved stellar populations of isolated early-type galaxies}},
	volume = {377},
	year = 2007}

@article{2005MNRAS.360..693R,
	adsurl = {https://ui.adsabs.harvard.edu/abs/2005MNRAS.360..693R},
	archiveprefix = {arXiv},
	author = {{Reda}, Fatma M. and {Forbes}, Duncan A. and {Hau}, George K.~T.},
	doi = {10.1111/j.1365-2966.2005.09058.x},
	eprint = {astro-ph/0505265},
	journal = {\mnras},
	month = jun,
	number = {2},
	pages = {693-702},
	primaryclass = {astro-ph},
	title = {{The fundamental plane of isolated early-type galaxies}},
	volume = {360},
	year = 2005}

@article{2004MNRAS.354..851R,
	adsurl = {https://ui.adsabs.harvard.edu/abs/2004MNRAS.354..851R},
	archiveprefix = {arXiv},
	author = {{Reda}, Fatma M. and {Forbes}, Duncan A. and {Beasley}, Michael A. and {O'Sullivan}, Ewan J. and {Goudfrooij}, Paul},
	doi = {10.1111/j.1365-2966.2004.08250.x},
	eprint = {astro-ph/0408424},
	journal = {\mnras},
	month = nov,
	number = {3},
	pages = {851-869},
	primaryclass = {astro-ph},
	title = {{The photometric properties of isolated early-type galaxies}},
	volume = {354},
	year = 2004}

@article{2015MNRAS.451.3427H,
	adsurl = {https://ui.adsabs.harvard.edu/abs/2015MNRAS.451.3427H},
	archiveprefix = {arXiv},
	author = {{Houghton}, R.~C.~W.},
	doi = {10.1093/mnras/stv1113},
	eprint = {1505.04788},
	journal = {\mnras},
	month = aug,
	number = {4},
	pages = {3427-3436},
	primaryclass = {astro-ph.GA},
	title = {{Revisiting the original morphology-density relation}},
	volume = {451},
	year = 2015}

@article{2003MNRAS.346..601G,
	adsurl = {https://ui.adsabs.harvard.edu/abs/2003MNRAS.346..601G},
	archiveprefix = {arXiv},
	author = {{Goto}, Tomotsugu and {Yamauchi}, Chisato and {Fujita}, Yutaka and {Okamura}, Sadanori and {Sekiguchi}, Maki and {Smail}, Ian and {Bernardi}, Mariangela and {Gomez}, Percy L.},
	doi = {10.1046/j.1365-2966.2003.07114.x},
	eprint = {astro-ph/0312043},
	journal = {\mnras},
	month = dec,
	number = {2},
	pages = {601-614},
	primaryclass = {astro-ph},
	title = {{The morphology-density relation in the Sloan Digital Sky Survey}},
	volume = {346},
	year = 2003}

@article{1984ApJ...281...95P,
	adsurl = {https://ui.adsabs.harvard.edu/abs/1984ApJ...281...95P},
	author = {{Postman}, M. and {Geller}, M.~J.},
	doi = {10.1086/162078},
	journal = {\apj},
	month = jun,
	pages = {95-99},
	title = {{The morphology-density relation - The group connection.}},
	volume = {281},
	year = 1984}

@article{2018ApJ...857..144H,
	adsurl = {https://ui.adsabs.harvard.edu/abs/2018ApJ...857..144H},
	archiveprefix = {arXiv},
	author = {{Hood}, Callie E. and {Kannappan}, Sheila J. and {Stark}, David V. and {Dell'Antonio}, Ian P. and {Moffett}, Amanda J. and {Eckert}, Kathleen D. and {Norris}, Mark A. and {Hendel}, David},
	doi = {10.3847/1538-4357/aab719},
	eid = {144},
	eprint = {1803.05447},
	journal = {\apj},
	month = apr,
	number = {2},
	pages = {144},
	primaryclass = {astro-ph.GA},
	title = {{The Origin of Faint Tidal Features around Galaxies in the RESOLVE Survey}},
	volume = {857},
	year = 2018}

@article{2024MNRAS.530.4422K,
	adsurl = {https://ui.adsabs.harvard.edu/abs/2024MNRAS.530.4422K},
	archiveprefix = {arXiv},
	author = {{Khalid}, A. and {Brough}, S. and {Martin}, G. and {Kimmig}, L.~C. and {Lagos}, C.~D.~P. and {Remus}, R. -S. and {Martinez-Lombilla}, C.},
	doi = {10.1093/mnras/stae1064},
	eprint = {2404.12436},
	journal = {\mnras},
	month = jun,
	number = {4},
	pages = {4422-4445},
	primaryclass = {astro-ph.GA},
	title = {{Characterizing tidal features around galaxies in cosmological simulations}},
	volume = {530},
	year = 2024}

@article{2024MNRAS.529..810R,
	adsurl = {https://ui.adsabs.harvard.edu/abs/2024MNRAS.529..810R},
	archiveprefix = {arXiv},
	author = {{Rutherford}, Tomas H. and {van de Sande}, Jesse and {Croom}, Scott M. and {Valenzuela}, Lucas M. and {Remus}, Rhea-Silvia and {D'Eugenio}, Francesco and {Vaughan}, Sam P. and {Zovaro}, Henry R.~M. and {Casura}, Sarah and {Barsanti}, Stefania and {Bland-Hawthorn}, Joss and {Brough}, Sarah and {Bryant}, Julia J. and {Goodwin}, Michael and {Lorente}, Nuria and {Oh}, Sree and {Ristea}, Andrei},
	doi = {10.1093/mnras/stae398},
	eprint = {2402.02728},
	journal = {\mnras},
	month = apr,
	number = {2},
	pages = {810-830},
	primaryclass = {astro-ph.GA},
	title = {{The SAMI Galaxy Survey: using tidal streams and shells to trace the dynamical evolution of massive galaxies}},
	volume = {529},
	year = 2024}

@article{2021A&C....3500469T,
	adsurl = {https://ui.adsabs.harvard.edu/abs/2021A&C....3500469T},
	archiveprefix = {arXiv},
	author = {{Tanoglidis}, D. and {{\'C}iprijanovi{\'c}}, A. and {Drlica-Wagner}, A.},
	doi = {10.1016/j.ascom.2021.100469},
	eid = {100469},
	eprint = {2011.12437},
	journal = {Astronomy and Computing},
	month = apr,
	pages = {100469},
	primaryclass = {astro-ph.GA},
	title = {{DeepShadows: Separating low surface brightness galaxies from artifacts using deep learning}},
	volume = {35},
	year = 2021}

@article{2021MNRAS.506.5030M,
	adsurl = {https://ui.adsabs.harvard.edu/abs/2021MNRAS.506.5030M},
	archiveprefix = {arXiv},
	author = {{Mart{\'\i}nez-Delgado}, David and {Rom{\'a}n}, Javier and {Erkal}, Denis and {Schirmer}, Mischa and {Roca-F{\`a}brega}, Santi and {Laine}, Seppo and {Donatiello}, Giuseppe and {Jimenez}, Manuel and {Malin}, David and {Carballo-Bello}, Julio A.},
	doi = {10.1093/mnras/stab1874},
	eprint = {2106.04548},
	journal = {\mnras},
	month = oct,
	number = {4},
	pages = {5030-5038},
	primaryclass = {astro-ph.GA},
	title = {{A feather on the hat: tracing the giant stellar stream around the Sombrero galaxy}},
	volume = {506},
	year = 2021}

@article{2008ApJ...689..936J,
	adsurl = {https://ui.adsabs.harvard.edu/abs/2008ApJ...689..936J},
	archiveprefix = {arXiv},
	author = {{Johnston}, Kathryn V. and {Bullock}, James S. and {Sharma}, Sanjib and {Font}, Andreea and {Robertson}, Brant E. and {Leitner}, Samuel N.},
	doi = {10.1086/592228},
	eprint = {0807.3911},
	journal = {\apj},
	month = dec,
	number = {2},
	pages = {936-957},
	primaryclass = {astro-ph},
	title = {{Tracing Galaxy Formation with Stellar Halos. II. Relating Substructure in Phase and Abundance Space to Accretion Histories}},
	volume = {689},
	year = 2008}

@article{2024ApJ...969..142C,
	adsurl = {https://ui.adsabs.harvard.edu/abs/2024ApJ...969..142C},
	archiveprefix = {arXiv},
	author = {{Chun}, Kyungwon and {Shin}, Jihye and {Ko}, Jongwan and {Smith}, Rory and {Yoo}, Jaewon},
	doi = {10.3847/1538-4357/ad4a52},
	eid = {142},
	eprint = {2405.08061},
	journal = {\apj},
	month = jul,
	number = {2},
	pages = {142},
	primaryclass = {astro-ph.GA},
	title = {{Formation Channels of Diffuse Lights in Groups and Clusters over Time}},
	volume = {969},
	year = 2024}

@article{2023ApJ...943..148C,
	adsurl = {https://ui.adsabs.harvard.edu/abs/2023ApJ...943..148C},
	archiveprefix = {arXiv},
	author = {{Chun}, Kyungwon and {Shin}, Jihye and {Smith}, Rory and {Ko}, Jongwan and {Yoo}, Jaewon},
	doi = {10.3847/1538-4357/aca890},
	eid = {148},
	eprint = {2212.02510},
	journal = {\apj},
	month = feb,
	number = {2},
	pages = {148},
	primaryclass = {astro-ph.GA},
	title = {{The Formation of the Brightest Cluster Galaxy and Intracluster Light in Cosmological N-body Simulations with the Galaxy Replacement Technique}},
	volume = {943},
	year = 2023}

@article{2022ApJ...925..103C,
	adsurl = {https://ui.adsabs.harvard.edu/abs/2022ApJ...925..103C},
	archiveprefix = {arXiv},
	author = {{Chun}, Kyungwon and {Shin}, Jihye and {Smith}, Rory and {Ko}, Jongwan and {Yoo}, Jaewon},
	doi = {10.3847/1538-4357/ac2cbe},
	eid = {103},
	eprint = {2110.09532},
	journal = {\apj},
	month = feb,
	number = {2},
	pages = {103},
	primaryclass = {astro-ph.GA},
	title = {{The Galaxy Replacement Technique (GRT): A New Approach to Study Tidal Stripping and Formation of Intracluster Light in a Cosmological Context}},
	volume = {925},
	year = 2022}

@article{2014A&A...566A..97J,
	adsurl = {https://ui.adsabs.harvard.edu/abs/2014A&A...566A..97J},
	archiveprefix = {arXiv},
	author = {{Ji}, Inchan and {Peirani}, S{\'e}bastien and {Yi}, Sukyoung K.},
	doi = {10.1051/0004-6361/201423530},
	eid = {A97},
	eprint = {1405.1807},
	journal = {\aap},
	month = jun,
	pages = {A97},
	primaryclass = {astro-ph.GA},
	title = {{Lifetime of merger features of equal-mass disk mergers}},
	volume = {566},
	year = 2014}

@article{2012AJ....144..128A,
	adsurl = {https://ui.adsabs.harvard.edu/abs/2012AJ....144..128A},
	archiveprefix = {arXiv},
	author = {{Adams}, Scott M. and {Zaritsky}, Dennis and {Sand}, David J. and {Graham}, Melissa L. and {Bildfell}, Chris and {Hoekstra}, Henk and {Pritchet}, Chris},
	doi = {10.1088/0004-6256/144/5/128},
	eid = {128},
	eprint = {1208.4843},
	journal = {\aj},
	month = nov,
	number = {5},
	pages = {128},
	primaryclass = {astro-ph.CO},
	title = {{The Environmental Dependence of the Incidence of Galactic Tidal Features}},
	volume = {144},
	year = 2012}

@article{2001AJ....121..808C,
	adsurl = {https://ui.adsabs.harvard.edu/abs/2001AJ....121..808C},
	archiveprefix = {arXiv},
	author = {{Colbert}, James W. and {Mulchaey}, John S. and {Zabludoff}, Ann I.},
	doi = {10.1086/318758},
	eprint = {astro-ph/0010534},
	journal = {\aj},
	month = feb,
	number = {2},
	pages = {808-819},
	primaryclass = {astro-ph},
	title = {{The Optical and Near-Infrared Morphologies of Isolated Early-Type Galaxies}},
	volume = {121},
	year = 2001}

@article{2024ApJ...965..158Y,
	adsurl = {https://ui.adsabs.harvard.edu/abs/2024ApJ...965..158Y},
	archiveprefix = {arXiv},
	author = {{Yoon}, Yongmin and {Ko}, Jongwan and {Chung}, Haeun and {Byun}, Woowon and {Chun}, Kyungwon},
	doi = {10.3847/1538-4357/ad34ad},
	eid = {158},
	eprint = {2404.03459},
	journal = {\apj},
	month = apr,
	number = {2},
	pages = {158},
	primaryclass = {astro-ph.GA},
	title = {{Shell-type Tidal Features Are More Frequently Detected in Slowly Rotating Early-type Galaxies than Stream- and Tail-type Features}},
	volume = {965},
	year = 2024}

@article{2023ApJ...946...41Y,
	adsurl = {https://ui.adsabs.harvard.edu/abs/2023ApJ...946...41Y},
	archiveprefix = {arXiv},
	author = {{Yoon}, Yongmin and {Ko}, Jongwan and {Kim}, Jae-Woo},
	doi = {10.3847/1538-4357/acbcc5},
	eid = {41},
	eprint = {2303.00559},
	journal = {\apj},
	month = mar,
	number = {1},
	pages = {41},
	primaryclass = {astro-ph.GA},
	title = {{Impact of Galaxy Mergers on Stellar Population Profiles of Early-type Galaxies}},
	volume = {946},
	year = 2023}

@article{2022ApJ...925..168Y,
	adsurl = {https://ui.adsabs.harvard.edu/abs/2022ApJ...925..168Y},
	archiveprefix = {arXiv},
	author = {{Yoon}, Yongmin and {Park}, Changbom and {Chung}, Haeun and {Lane}, Richard R.},
	doi = {10.3847/1538-4357/ac415d},
	eid = {168},
	eprint = {2112.13703},
	journal = {\apj},
	month = feb,
	number = {2},
	pages = {168},
	primaryclass = {astro-ph.GA},
	title = {{Evidence for Impact of Galaxy Mergers on Stellar Kinematics of Early-type Galaxies}},
	volume = {925},
	year = 2022}

@article{2020ApJ...905..154Y,
	adsurl = {https://ui.adsabs.harvard.edu/abs/2020ApJ...905..154Y},
	archiveprefix = {arXiv},
	author = {{Yoon}, Yongmin and {Lim}, Gu},
	doi = {10.3847/1538-4357/abc621},
	eid = {154},
	eprint = {2011.14689},
	journal = {\apj},
	month = dec,
	number = {2},
	pages = {154},
	primaryclass = {astro-ph.GA},
	title = {{Frequency of Tidal Features Correlates with Age and Internal Structure of Early-type Galaxies}},
	volume = {905},
	year = 2020}

@article{2022ApJ...935..167A,
	adsurl = {https://ui.adsabs.harvard.edu/abs/2022ApJ...935..167A},
	archiveprefix = {arXiv},
	author = {{Astropy Collaboration} and {Price-Whelan}, Adrian M. and {Lim}, Pey Lian and {Earl}, Nicholas and {Starkman}, Nathaniel and {Bradley}, Larry and {Shupe}, David L. and {Patil}, Aarya A. and {Corrales}, Lia and {Brasseur}, C.~E. and {N{\"o}the}, Maximilian and {Donath}, Axel and {Tollerud}, Erik and {Morris}, Brett M. and {Ginsburg}, Adam and {Vaher}, Eero and {Weaver}, Benjamin A. and {Tocknell}, James and {Jamieson}, William and {van Kerkwijk}, Marten H. and {Robitaille}, Thomas P. and {Merry}, Bruce and {Bachetti}, Matteo and {G{\"u}nther}, H. Moritz and {Aldcroft}, Thomas L. and {Alvarado-Montes}, Jaime A. and {Archibald}, Anne M. and {B{\'o}di}, Attila and {Bapat}, Shreyas and {Barentsen}, Geert and {Baz{\'a}n}, Juanjo and {Biswas}, Manish and {Boquien}, M{\'e}d{\'e}ric and {Burke}, D.~J. and {Cara}, Daria and {Cara}, Mihai and {Conroy}, Kyle E. and {Conseil}, Simon and {Craig}, Matthew W. and {Cross}, Robert M. and {Cruz}, Kelle L. and {D'Eugenio}, Francesco and {Dencheva}, Nadia and {Devillepoix}, Hadrien A.~R. and {Dietrich}, J{\"o}rg P. and {Eigenbrot}, Arthur Davis and {Erben}, Thomas and {Ferreira}, Leonardo and {Foreman-Mackey}, Daniel and {Fox}, Ryan and {Freij}, Nabil and {Garg}, Suyog and {Geda}, Robel and {Glattly}, Lauren and {Gondhalekar}, Yash and {Gordon}, Karl D. and {Grant}, David and {Greenfield}, Perry and {Groener}, Austen M. and {Guest}, Steve and {Gurovich}, Sebastian and {Handberg}, Rasmus and {Hart}, Akeem and {Hatfield-Dodds}, Zac and {Homeier}, Derek and {Hosseinzadeh}, Griffin and {Jenness}, Tim and {Jones}, Craig K. and {Joseph}, Prajwel and {Kalmbach}, J. Bryce and {Karamehmetoglu}, Emir and {Ka{\l}uszy{\'n}ski}, Miko{\l}aj and {Kelley}, Michael S.~P. and {Kern}, Nicholas and {Kerzendorf}, Wolfgang E. and {Koch}, Eric W. and {Kulumani}, Shankar and {Lee}, Antony and {Ly}, Chun and {Ma}, Zhiyuan and {MacBride}, Conor and {Maljaars}, Jakob M. and {Muna}, Demitri and {Murphy}, N.~A. and {Norman}, Henrik and {O'Steen}, Richard and {Oman}, Kyle A. and {Pacifici}, Camilla and {Pascual}, Sergio and {Pascual-Granado}, J. and {Patil}, Rohit R. and {Perren}, Gabriel I. and {Pickering}, Timothy E. and {Rastogi}, Tanuj and {Roulston}, Benjamin R. and {Ryan}, Daniel F. and {Rykoff}, Eli S. and {Sabater}, Jose and {Sakurikar}, Parikshit and {Salgado}, Jes{\'u}s and {Sanghi}, Aniket and {Saunders}, Nicholas and {Savchenko}, Volodymyr and {Schwardt}, Ludwig and {Seifert-Eckert}, Michael and {Shih}, Albert Y. and {Jain}, Anany Shrey and {Shukla}, Gyanendra and {Sick}, Jonathan and {Simpson}, Chris and {Singanamalla}, Sudheesh and {Singer}, Leo P. and {Singhal}, Jaladh and {Sinha}, Manodeep and {Sip{\H{o}}cz}, Brigitta M. and {Spitler}, Lee R. and {Stansby}, David and {Streicher}, Ole and {{\v{S}}umak}, Jani and {Swinbank}, John D. and {Taranu}, Dan S. and {Tewary}, Nikita and {Tremblay}, Grant R. and {de Val-Borro}, Miguel and {Van Kooten}, Samuel J. and {Vasovi{\'c}}, Zlatan and {Verma}, Shresth and {de Miranda Cardoso}, Jos{\'e} Vin{\'\i}cius and {Williams}, Peter K.~G. and {Wilson}, Tom J. and {Winkel}, Benjamin and {Wood-Vasey}, W.~M. and {Xue}, Rui and {Yoachim}, Peter and {Zhang}, Chen and {Zonca}, Andrea and {Astropy Project Contributors}},
	doi = {10.3847/1538-4357/ac7c74},
	eid = {167},
	eprint = {2206.14220},
	journal = {\apj},
	month = aug,
	number = {2},
	pages = {167},
	primaryclass = {astro-ph.IM},
	title = {{The Astropy Project: Sustaining and Growing a Community-oriented Open-source Project and the Latest Major Release (v5.0) of the Core Package}},
	volume = {935},
	year = 2022}

@software{larry_bradley_2022_6825092,
	author = {Larry Bradley and Brigitta Sip{\H o}cz and Thomas Robitaille and Erik Tollerud and Z{\'e} Vin{\'\i}cius and Christoph Deil and Kyle Barbary and Tom J Wilson and Ivo Busko and Axel Donath and Hans Moritz G{\"u}nther and Mihai Cara and P. L. Lim and Sebastian Me{\ss}linger and Simon Conseil and Azalee Bostroem and Michael Droettboom and E. M. Bray and Lars Andersen Bratholm and Geert Barentsen and Matt Craig and Shivangee Rathi and Sergio Pascual and Gabriel Perren and Iskren Y. Georgiev and Miguel de Val-Borro and Wolfgang Kerzendorf and Yoonsoo P. Bach and Bruno Quint and Harrison Souchereau},
	doi = {10.5281/zenodo.6825092},
	month = jul,
	publisher = {Zenodo},
	title = {astropy/photutils: 1.5.0},
	url = {https://doi.org/10.5281/zenodo.6825092},
	version = {1.5.0},
	year = 2022}

@article{2016MNRAS.456.1359F,
	adsurl = {https://ui.adsabs.harvard.edu/abs/2016MNRAS.456.1359F},
	archiveprefix = {arXiv},
	author = {{Fliri}, J{\"u}rgen and {Trujillo}, Ignacio},
	doi = {10.1093/mnras/stv2686},
	eprint = {1603.04474},
	journal = {\mnras},
	month = feb,
	number = {2},
	pages = {1359-1373},
	primaryclass = {astro-ph.GA},
	title = {{The IAC Stripe 82 Legacy Project: a wide-area survey for faint surface brightness astronomy}},
	volume = {456},
	year = 2016}

@article{2014ApJS..213...12J,
	adsurl = {https://ui.adsabs.harvard.edu/abs/2014ApJS..213...12J},
	archiveprefix = {arXiv},
	author = {{Jiang}, Linhua and {Fan}, Xiaohui and {Bian}, Fuyan and {McGreer}, Ian D. and {Strauss}, Michael A. and {Annis}, James and {Buck}, Zo{\"e} and {Green}, Richard and {Hodge}, Jacqueline A. and {Myers}, Adam D. and {Rafiee}, Alireza and {Richards}, Gordon},
	doi = {10.1088/0067-0049/213/1/12},
	eid = {12},
	eprint = {1405.7382},
	journal = {\apjs},
	month = jul,
	number = {1},
	pages = {12},
	primaryclass = {astro-ph.GA},
	title = {{The Sloan Digital Sky Survey Stripe 82 Imaging Data: Depth-optimized Co-adds over 300 deg$^{2}$ in Five Filters}},
	volume = {213},
	year = 2014}

@article{2019AJ....157..168D,
	adsurl = {https://ui.adsabs.harvard.edu/abs/2019AJ....157..168D},
	archiveprefix = {arXiv},
	author = {{Dey}, Arjun and {Schlegel}, David J. and {Lang}, Dustin and {Blum}, Robert and {Burleigh}, Kaylan and {Fan}, Xiaohui and {Findlay}, Joseph R. and {Finkbeiner}, Doug and {Herrera}, David and {Juneau}, St{\'e}phanie and {Landriau}, Martin and {Levi}, Michael and {McGreer}, Ian and {Meisner}, Aaron and {Myers}, Adam D. and {Moustakas}, John and {Nugent}, Peter and {Patej}, Anna and {Schlafly}, Edward F. and {Walker}, Alistair R. and {Valdes}, Francisco and {Weaver}, Benjamin A. and {Y{\`e}che}, Christophe and {Zou}, Hu and {Zhou}, Xu and {Abareshi}, Behzad and {Abbott}, T.~M.~C. and {Abolfathi}, Bela and {Aguilera}, C. and {Alam}, Shadab and {Allen}, Lori and {Alvarez}, A. and {Annis}, James and {Ansarinejad}, Behzad and {Aubert}, Marie and {Beechert}, Jacqueline and {Bell}, Eric F. and {BenZvi}, Segev Y. and {Beutler}, Florian and {Bielby}, Richard M. and {Bolton}, Adam S. and {Brice{\~n}o}, C{\'e}sar and {Buckley-Geer}, Elizabeth J. and {Butler}, Karen and {Calamida}, Annalisa and {Carlberg}, Raymond G. and {Carter}, Paul and {Casas}, Ricard and {Castander}, Francisco J. and {Choi}, Yumi and {Comparat}, Johan and {Cukanovaite}, Elena and {Delubac}, Timoth{\'e}e and {DeVries}, Kaitlin and {Dey}, Sharmila and {Dhungana}, Govinda and {Dickinson}, Mark and {Ding}, Zhejie and {Donaldson}, John B. and {Duan}, Yutong and {Duckworth}, Christopher J. and {Eftekharzadeh}, Sarah and {Eisenstein}, Daniel J. and {Etourneau}, Thomas and {Fagrelius}, Parker A. and {Farihi}, Jay and {Fitzpatrick}, Mike and {Font-Ribera}, Andreu and {Fulmer}, Leah and {G{\"a}nsicke}, Boris T. and {Gaztanaga}, Enrique and {George}, Koshy and {Gerdes}, David W. and {Gontcho}, Satya Gontcho A. and {Gorgoni}, Claudio and {Green}, Gregory and {Guy}, Julien and {Harmer}, Diane and {Hernandez}, M. and {Honscheid}, Klaus and {Huang}, Lijuan Wendy and {James}, David J. and {Jannuzi}, Buell T. and {Jiang}, Linhua and {Joyce}, Richard and {Karcher}, Armin and {Karkar}, Sonia and {Kehoe}, Robert and {Kneib}, Jean-Paul and {Kueter-Young}, Andrea and {Lan}, Ting-Wen and {Lauer}, Tod R. and {Le Guillou}, Laurent and {Le Van Suu}, Auguste and {Lee}, Jae Hyeon and {Lesser}, Michael and {Perreault Levasseur}, Laurence and {Li}, Ting S. and {Mann}, Justin L. and {Marshall}, Robert and {Mart{\'\i}nez-V{\'a}zquez}, C.~E. and {Martini}, Paul and {du Mas des Bourboux}, H{\'e}lion and {McManus}, Sean and {Meier}, Tobias Gabriel and {M{\'e}nard}, Brice and {Metcalfe}, Nigel and {Mu{\~n}oz-Guti{\'e}rrez}, Andrea and {Najita}, Joan and {Napier}, Kevin and {Narayan}, Gautham and {Newman}, Jeffrey A. and {Nie}, Jundan and {Nord}, Brian and {Norman}, Dara J. and {Olsen}, Knut A.~G. and {Paat}, Anthony and {Palanque-Delabrouille}, Nathalie and {Peng}, Xiyan and {Poppett}, Claire L. and {Poremba}, Megan R. and {Prakash}, Abhishek and {Rabinowitz}, David and {Raichoor}, Anand and {Rezaie}, Mehdi and {Robertson}, A.~N. and {Roe}, Natalie A. and {Ross}, Ashley J. and {Ross}, Nicholas P. and {Rudnick}, Gregory and {Safonova}, Sasha and {Saha}, Abhijit and {S{\'a}nchez}, F. Javier and {Savary}, Elodie and {Schweiker}, Heidi and {Scott}, Adam and {Seo}, Hee-Jong and {Shan}, Huanyuan and {Silva}, David R. and {Slepian}, Zachary and {Soto}, Christian and {Sprayberry}, David and {Staten}, Ryan and {Stillman}, Coley M. and {Stupak}, Robert J. and {Summers}, David L. and {Sien Tie}, Suk and {Tirado}, H. and {Vargas-Maga{\~n}a}, Mariana and {Vivas}, A. Katherina and {Wechsler}, Risa H. and {Williams}, Doug and {Yang}, Jinyi and {Yang}, Qian and {Yapici}, Tolga and {Zaritsky}, Dennis and {Zenteno}, A. and {Zhang}, Kai and {Zhang}, Tianmeng and {Zhou}, Rongpu and {Zhou}, Zhimin},
	doi = {10.3847/1538-3881/ab089d},
	eid = {168},
	eprint = {1804.08657},
	journal = {\aj},
	month = may,
	number = {5},
	pages = {168},
	primaryclass = {astro-ph.IM},
	title = {{Overview of the DESI Legacy Imaging Surveys}},
	volume = {157},
	year = 2019}

@article{2016MNRAS.458.2371R,
	adsurl = {https://ui.adsabs.harvard.edu/abs/2016MNRAS.458.2371R},
	archiveprefix = {arXiv},
	author = {{Rodriguez-Gomez}, Vicente and {Pillepich}, Annalisa and {Sales}, Laura V. and {Genel}, Shy and {Vogelsberger}, Mark and {Zhu}, Qirong and {Wellons}, Sarah and {Nelson}, Dylan and {Torrey}, Paul and {Springel}, Volker and {Ma}, Chung-Pei and {Hernquist}, Lars},
	doi = {10.1093/mnras/stw456},
	eprint = {1511.08804},
	journal = {\mnras},
	month = may,
	number = {3},
	pages = {2371-2390},
	primaryclass = {astro-ph.GA},
	title = {{The stellar mass assembly of galaxies in the Illustris simulation: growth by mergers and the spatial distribution of accreted stars}},
	volume = {458},
	year = 2016}

@article{2013MNRAS.434.3348C,
	adsurl = {https://ui.adsabs.harvard.edu/abs/2013MNRAS.434.3348C},
	archiveprefix = {arXiv},
	author = {{Cooper}, Andrew P. and {D'Souza}, Richard and {Kauffmann}, Guinevere and {Wang}, Jing and {Boylan-Kolchin}, Michael and {Guo}, Qi and {Frenk}, Carlos S. and {White}, Simon D.~M.},
	doi = {10.1093/mnras/stt1245},
	eprint = {1303.6283},
	journal = {\mnras},
	month = oct,
	number = {4},
	pages = {3348-3367},
	primaryclass = {astro-ph.CO},
	title = {{Galactic accretion and the outer structure of galaxies in the CDM model}},
	volume = {434},
	year = 2013}

@article{2010ApJ...725.2312O,
	adsurl = {https://ui.adsabs.harvard.edu/abs/2010ApJ...725.2312O},
	archiveprefix = {arXiv},
	author = {{Oser}, Ludwig and {Ostriker}, Jeremiah P. and {Naab}, Thorsten and {Johansson}, Peter H. and {Burkert}, Andreas},
	doi = {10.1088/0004-637X/725/2/2312},
	eprint = {1010.1381},
	journal = {\apj},
	month = dec,
	number = {2},
	pages = {2312-2323},
	primaryclass = {astro-ph.CO},
	title = {{The Two Phases of Galaxy Formation}},
	volume = {725},
	year = 2010}

@article{2006ApJ...647L..99D,
	adsurl = {https://ui.adsabs.harvard.edu/abs/2006ApJ...647L..99D},
	author = {{di Serego Alighieri}, Sperello and {Lanzoni}, Barbara and {J{\o}rgensen}, Inger},
	doi = {10.1086/507582},
	journal = {\apjl},
	month = aug,
	number = {2},
	pages = {L99-L102},
	title = {{The Influence of Mass and Environment on the Evolution of Early-Type Galaxies}},
	volume = {647},
	year = 2006}

@article{2005ApJ...621..673T,
	adsurl = {https://ui.adsabs.harvard.edu/abs/2005ApJ...621..673T},
	archiveprefix = {arXiv},
	author = {{Thomas}, Daniel and {Maraston}, Claudia and {Bender}, Ralf and {Mendes de Oliveira}, Claudia},
	doi = {10.1086/426932},
	eprint = {astro-ph/0410209},
	journal = {\apj},
	month = mar,
	number = {2},
	pages = {673-694},
	primaryclass = {astro-ph},
	title = {{The Epochs of Early-Type Galaxy Formation as a Function of Environment}},
	volume = {621},
	year = 2005}

@article{1989ARA&A..27..235K,
	adsurl = {https://ui.adsabs.harvard.edu/abs/1989ARA&A..27..235K},
	author = {{Kormendy}, John and {Djorgovski}, S.},
	doi = {10.1146/annurev.aa.27.090189.001315},
	journal = {\araa},
	month = jan,
	pages = {235-277},
	title = {{Surface photometry and the structure of elliptical galaxies.}},
	volume = {27},
	year = 1989}

@article{1976ApJ...204..668F,
	adsurl = {https://ui.adsabs.harvard.edu/abs/1976ApJ...204..668F},
	author = {{Faber}, S.~M. and {Jackson}, R.~E.},
	doi = {10.1086/154215},
	journal = {\apj},
	month = mar,
	pages = {668-683},
	title = {{Velocity dispersions and mass-to-light ratios for elliptical galaxies.}},
	volume = {204},
	year = 1976}

@article{2009ApJ...699L.178N,
	adsurl = {https://ui.adsabs.harvard.edu/abs/2009ApJ...699L.178N},
	archiveprefix = {arXiv},
	author = {{Naab}, Thorsten and {Johansson}, Peter H. and {Ostriker}, Jeremiah P.},
	doi = {10.1088/0004-637X/699/2/L178},
	eprint = {0903.1636},
	journal = {\apjl},
	month = jul,
	number = {2},
	pages = {L178-L182},
	primaryclass = {astro-ph.CO},
	title = {{Minor Mergers and the Size Evolution of Elliptical Galaxies}},
	volume = {699},
	year = 2009}

@article{2010ApJ...709.1018V,
	adsurl = {https://ui.adsabs.harvard.edu/abs/2010ApJ...709.1018V},
	archiveprefix = {arXiv},
	author = {{van Dokkum}, Pieter G. and {Whitaker}, Katherine E. and {Brammer}, Gabriel and {Franx}, Marijn and {Kriek}, Mariska and {Labb{\'e}}, Ivo and {Marchesini}, Danilo and {Quadri}, Ryan and {Bezanson}, Rachel and {Illingworth}, Garth D. and {Muzzin}, Adam and {Rudnick}, Gregory and {Tal}, Tomer and {Wake}, David},
	doi = {10.1088/0004-637X/709/2/1018},
	eprint = {0912.0514},
	journal = {\apj},
	month = feb,
	number = {2},
	pages = {1018-1041},
	primaryclass = {astro-ph.CO},
	title = {{The Growth of Massive Galaxies Since z = 2}},
	volume = {709},
	year = 2010}

@article{1992ApJ...399..462B,
	adsurl = {https://ui.adsabs.harvard.edu/abs/1992ApJ...399..462B},
	author = {{Bender}, Ralf and {Burstein}, David and {Faber}, S.~M.},
	doi = {10.1086/171940},
	journal = {\apj},
	month = nov,
	pages = {462},
	title = {{Dynamically Hot Galaxies. I. Structural Properties}},
	volume = {399},
	year = 1992}

@inproceedings{1977egsp.conf..401T,
	adsurl = {https://ui.adsabs.harvard.edu/abs/1977egsp.conf..401T},
	author = {{Toomre}, Alar},
	booktitle = {Evolution of Galaxies and Stellar Populations},
	editor = {{Tinsley}, Beatrice M. and {Larson}, D. Campbell, Richard B. Gehret},
	month = jan,
	pages = {401},
	title = {{Mergers and Some Consequences}},
	year = 1977}

@article{1980ApJ...236..351D,
	adsurl = {https://ui.adsabs.harvard.edu/abs/1980ApJ...236..351D},
	author = {{Dressler}, A.},
	doi = {10.1086/157753},
	journal = {\apj},
	month = mar,
	pages = {351-365},
	title = {{Galaxy morphology in rich clusters: implications for the formation and evolution of galaxies.}},
	volume = {236},
	year = 1980}

@article{2014MNRAS.443.1433D,
	adsurl = {https://ui.adsabs.harvard.edu/abs/2014MNRAS.443.1433D},
	archiveprefix = {arXiv},
	author = {{D'Souza}, Richard and {Kauffman}, Guinevere and {Wang}, Jing and {Vegetti}, Simona},
	doi = {10.1093/mnras/stu1194},
	eprint = {1404.2123},
	journal = {\mnras},
	month = sep,
	number = {2},
	pages = {1433-1450},
	primaryclass = {astro-ph.GA},
	title = {{Parametrizing the stellar haloes of galaxies}},
	volume = {443},
	year = 2014}

@article{2013MNRAS.432.3391T,
	adsurl = {https://ui.adsabs.harvard.edu/abs/2013MNRAS.432.3391T},
	archiveprefix = {arXiv},
	author = {{Tissera}, Patricia B. and {Scannapieco}, Cecilia and {Beers}, Timothy C. and {Carollo}, Daniela},
	doi = {10.1093/mnras/stt691},
	eprint = {1301.1301},
	journal = {\mnras},
	month = jul,
	number = {4},
	pages = {3391-3400},
	primaryclass = {astro-ph.GA},
	title = {{Stellar haloes of simulated Milky-Way-like galaxies: chemical and kinematic properties}},
	volume = {432},
	year = 2013}

@article{2022ApJ...932...44G,
	adsurl = {https://ui.adsabs.harvard.edu/abs/2022ApJ...932...44G},
	archiveprefix = {arXiv},
	author = {{Gilhuly}, Colleen and {Merritt}, Allison and {Abraham}, Roberto and {Danieli}, Shany and {Lokhorst}, Deborah and {Liu}, Qing and {van Dokkum}, Pieter and {Conroy}, Charlie and {Greco}, Johnny},
	doi = {10.3847/1538-4357/ac6750},
	eid = {44},
	eprint = {2204.06596},
	journal = {\apj},
	month = jun,
	number = {1},
	pages = {44},
	primaryclass = {astro-ph.GA},
	title = {{Stellar Halos from the The Dragonfly Edge-on Galaxies Survey}},
	volume = {932},
	year = 2022}

@article{2020MNRAS.495.4570M,
	adsurl = {https://ui.adsabs.harvard.edu/abs/2020MNRAS.495.4570M},
	archiveprefix = {arXiv},
	author = {{Merritt}, Allison and {Pillepich}, Annalisa and {van Dokkum}, Pieter and {Nelson}, Dylan and {Hernquist}, Lars and {Marinacci}, Federico and {Vogelsberger}, Mark},
	doi = {10.1093/mnras/staa1164},
	eprint = {2004.11402},
	journal = {\mnras},
	month = jul,
	number = {4},
	pages = {4570-4604},
	primaryclass = {astro-ph.GA},
	title = {{A missing outskirts problem? Comparisons between stellar haloes in the Dragonfly Nearby Galaxies Survey and the TNG100 simulation}},
	volume = {495},
	year = 2020}

@article{2014MNRAS.444..237P,
	adsurl = {https://ui.adsabs.harvard.edu/abs/2014MNRAS.444..237P},
	archiveprefix = {arXiv},
	author = {{Pillepich}, Annalisa and {Vogelsberger}, Mark and {Deason}, Alis and {Rodriguez-Gomez}, Vicente and {Genel}, Shy and {Nelson}, Dylan and {Torrey}, Paul and {Sales}, Laura V. and {Marinacci}, Federico and {Springel}, Volker and {Sijacki}, Debora and {Hernquist}, Lars},
	doi = {10.1093/mnras/stu1408},
	eprint = {1406.1174},
	journal = {\mnras},
	month = oct,
	number = {1},
	pages = {237-249},
	primaryclass = {astro-ph.GA},
	title = {{Halo mass and assembly history exposed in the faint outskirts: the stellar and dark matter haloes of Illustris galaxies}},
	volume = {444},
	year = 2014}

@article{2018ApJ...869...12S,
	adsurl = {https://ui.adsabs.harvard.edu/abs/2018ApJ...869...12S},
	archiveprefix = {arXiv},
	author = {{Sanderson}, Robyn E. and {Garrison-Kimmel}, Shea and {Wetzel}, Andrew and {Keung Chan}, Tsang and {Hopkins}, Philip F. and {Kere{\v{s}}}, Du{\v{s}}an and {Escala}, Ivanna and {Faucher-Gigu{\`e}re}, Claude-Andr{\'e} and {Ma}, Xiangcheng},
	doi = {10.3847/1538-4357/aaeb33},
	eid = {12},
	eprint = {1712.05808},
	journal = {\apj},
	month = dec,
	number = {1},
	pages = {12},
	primaryclass = {astro-ph.GA},
	title = {{Reconciling Observed and Simulated Stellar Halo Masses}},
	volume = {869},
	year = 2018}

@article{2016ApJ...830...62M,
	adsurl = {https://ui.adsabs.harvard.edu/abs/2016ApJ...830...62M},
	archiveprefix = {arXiv},
	author = {{Merritt}, Allison and {van Dokkum}, Pieter and {Abraham}, Roberto and {Zhang}, Jielai},
	doi = {10.3847/0004-637X/830/2/62},
	eid = {62},
	eprint = {1606.08847},
	journal = {\apj},
	month = oct,
	number = {2},
	pages = {62},
	primaryclass = {astro-ph.GA},
	title = {{The Dragonfly nearby Galaxies Survey. I. Substantial Variation in the Diffuse Stellar Halos around Spiral Galaxies}},
	volume = {830},
	year = 2016}

@article{2024AJ....167...57T,
	adsurl = {https://ui.adsabs.harvard.edu/abs/2024AJ....167...57T},
	archiveprefix = {arXiv},
	author = {{Tau}, Elisa A. and {Vivas}, A. Katherina and {Mart{\'\i}nez-V{\'a}zquez}, Clara E.},
	doi = {10.3847/1538-3881/ad1509},
	eid = {57},
	eprint = {2312.07279},
	journal = {\aj},
	month = feb,
	number = {2},
	pages = {57},
	primaryclass = {astro-ph.GA},
	title = {{Extended Stellar Populations in Ultrafaint Dwarf Galaxies}},
	volume = {167},
	year = 2024}

@article{2023MNRAS.519.1349W,
	adsurl = {https://ui.adsabs.harvard.edu/abs/2023MNRAS.519.1349W},
	archiveprefix = {arXiv},
	author = {{Waller}, Fletcher and {Venn}, Kim A. and {Sestito}, Federico and {Jensen}, Jaclyn and {Kielty}, Collin L. and {Borukhovetskaya}, Asya and {Hayes}, Christian and {McConnachie}, Alan W. and {Navarro}, Julio F.},
	doi = {10.1093/mnras/stac3563},
	eprint = {2208.07948},
	journal = {\mnras},
	month = feb,
	number = {1},
	pages = {1349-1365},
	primaryclass = {astro-ph.GA},
	title = {{The Cosmic Hunt for members in the outskirts of ultra-faint dwarf galaxies: Ursa Major I, Coma Berenices, and Bo{\"o}tes I}},
	volume = {519},
	year = 2023}

@article{2024MNRAS.527.4209J,
	adsurl = {https://ui.adsabs.harvard.edu/abs/2024MNRAS.527.4209J},
	archiveprefix = {arXiv},
	author = {{Jensen}, Jaclyn and {Hayes}, Christian R. and {Sestito}, Federico and {McConnachie}, Alan W. and {Waller}, Fletcher and {Smith}, Simon E.~T. and {Navarro}, Julio and {Venn}, Kim A.},
	doi = {10.1093/mnras/stad3322},
	eprint = {2308.07394},
	journal = {\mnras},
	month = jan,
	number = {2},
	pages = {4209-4233},
	primaryclass = {astro-ph.GA},
	title = {{Small-scale stellar haloes: detecting low surface brightness features in the outskirts of Milky Way dwarf satellites}},
	volume = {527},
	year = 2024}

@article{2023MNRAS.523..123S,
	adsurl = {https://ui.adsabs.harvard.edu/abs/2023MNRAS.523..123S},
	archiveprefix = {arXiv},
	author = {{Sestito}, Federico and {Roediger}, Joel and {Navarro}, Julio F. and {Jensen}, Jaclyn and {Venn}, Kim A. and {Smith}, Simon E.~T. and {Hayes}, Christian and {McConnachie}, Alan W.},
	doi = {10.1093/mnras/stad1417},
	eprint = {2304.13048},
	journal = {\mnras},
	month = jul,
	number = {1},
	pages = {123-131},
	primaryclass = {astro-ph.GA},
	title = {{Stars on the edge: Galactic tides and the outskirts of the Sculptor dwarf spheroidal}},
	volume = {523},
	year = 2023}

@article{2023MNRAS.525.2875S,
	adsurl = {https://ui.adsabs.harvard.edu/abs/2023MNRAS.525.2875S},
	archiveprefix = {arXiv},
	author = {{Sestito}, Federico and {Zaremba}, Daria and {Venn}, Kim A. and {D'Aoust}, Lina and {Hayes}, Christian and {Jensen}, Jaclyn and {Navarro}, Julio F. and {Jablonka}, Pascale and {Fern{\'a}ndez-Alvar}, Emma and {Glover}, Jennifer and {McConnachie}, Alan W. and {Chen{\'e}}, Andr{\'e}-Nicolas},
	doi = {10.1093/mnras/stad2427},
	eprint = {2301.13214},
	journal = {\mnras},
	month = oct,
	number = {2},
	pages = {2875-2890},
	primaryclass = {astro-ph.GA},
	title = {{The extended 'stellar halo' of the Ursa Minor dwarf galaxy}},
	volume = {525},
	year = 2023}

@article{1996A&AS..117..393B,
	adsurl = {https://ui.adsabs.harvard.edu/abs/1996A&AS..117..393B},
	author = {{Bertin}, E. and {Arnouts}, S.},
	doi = {10.1051/aas:1996164},
	journal = {\aaps},
	month = {Jun},
	pages = {393-404},
	title = {{SExtractor: Software for source extraction.}},
	volume = {117},
	year = {1996}}

@article{2020NatMe..17..261V,
	adsurl = {https://ui.adsabs.harvard.edu/abs/2020NatMe..17..261V},
	archiveprefix = {arXiv},
	author = {{Virtanen}, Pauli and {Gommers}, Ralf and {Oliphant}, Travis E. and {Haberland}, Matt and {Reddy}, Tyler and {Cournapeau}, David and {Burovski}, Evgeni and {Peterson}, Pearu and {Weckesser}, Warren and {Bright}, Jonathan and {van der Walt}, St{\'e}fan J. and {Brett}, Matthew and {Wilson}, Joshua and {Millman}, K. Jarrod and {Mayorov}, Nikolay and {Nelson}, Andrew R.~J. and {Jones}, Eric and {Kern}, Robert and {Larson}, Eric and {Carey}, C.~J. and {Polat}, {\.I}lhan and {Feng}, Yu and {Moore}, Eric W. and {VanderPlas}, Jake and {Laxalde}, Denis and {Perktold}, Josef and {Cimrman}, Robert and {Henriksen}, Ian and {Quintero}, E.~A. and {Harris}, Charles R. and {Archibald}, Anne M. and {Ribeiro}, Ant{\^o}nio H. and {Pedregosa}, Fabian and {van Mulbregt}, Paul and {SciPy 1. 0 Contributors}},
	doi = {10.1038/s41592-019-0686-2},
	eprint = {1907.10121},
	journal = {Nature Methods},
	month = feb,
	pages = {261-272},
	primaryclass = {cs.MS},
	title = {{SciPy 1.0: fundamental algorithms for scientific computing in Python}},
	volume = {17},
	year = 2020}

@article{2018AJ....156..123A,
	adsurl = {https://ui.adsabs.harvard.edu/abs/2018AJ....156..123A},
	archiveprefix = {arXiv},
	author = {{Astropy Collaboration} and {Price-Whelan}, A.~M. and {Sip{\H{o}}cz}, B.~M. and {G{\"u}nther}, H.~M. and {Lim}, P.~L. and {Crawford}, S.~M. and {Conseil}, S. and {Shupe}, D.~L. and {Craig}, M.~W. and {Dencheva}, N. and {Ginsburg}, A. and {VanderPlas}, J.~T. and {Bradley}, L.~D. and {P{\'e}rez-Su{\'a}rez}, D. and {de Val-Borro}, M. and {Aldcroft}, T.~L. and {Cruz}, K.~L. and {Robitaille}, T.~P. and {Tollerud}, E.~J. and {Ardelean}, C. and {Babej}, T. and {Bach}, Y.~P. and {Bachetti}, M. and {Bakanov}, A.~V. and {Bamford}, S.~P. and {Barentsen}, G. and {Barmby}, P. and {Baumbach}, A. and {Berry}, K.~L. and {Biscani}, F. and {Boquien}, M. and {Bostroem}, K.~A. and {Bouma}, L.~G. and {Brammer}, G.~B. and {Bray}, E.~M. and {Breytenbach}, H. and {Buddelmeijer}, H. and {Burke}, D.~J. and {Calderone}, G. and {Cano Rodr{\'\i}guez}, J.~L. and {Cara}, M. and {Cardoso}, J.~V.~M. and {Cheedella}, S. and {Copin}, Y. and {Corrales}, L. and {Crichton}, D. and {D'Avella}, D. and {Deil}, C. and {Depagne}, {\'E}. and {Dietrich}, J.~P. and {Donath}, A. and {Droettboom}, M. and {Earl}, N. and {Erben}, T. and {Fabbro}, S. and {Ferreira}, L.~A. and {Finethy}, T. and {Fox}, R.~T. and {Garrison}, L.~H. and {Gibbons}, S.~L.~J. and {Goldstein}, D.~A. and {Gommers}, R. and {Greco}, J.~P. and {Greenfield}, P. and {Groener}, A.~M. and {Grollier}, F. and {Hagen}, A. and {Hirst}, P. and {Homeier}, D. and {Horton}, A.~J. and {Hosseinzadeh}, G. and {Hu}, L. and {Hunkeler}, J.~S. and {Ivezi{\'c}}, {\v{Z}}. and {Jain}, A. and {Jenness}, T. and {Kanarek}, G. and {Kendrew}, S. and {Kern}, N.~S. and {Kerzendorf}, W.~E. and {Khvalko}, A. and {King}, J. and {Kirkby}, D. and {Kulkarni}, A.~M. and {Kumar}, A. and {Lee}, A. and {Lenz}, D. and {Littlefair}, S.~P. and {Ma}, Z. and {Macleod}, D.~M. and {Mastropietro}, M. and {McCully}, C. and {Montagnac}, S. and {Morris}, B.~M. and {Mueller}, M. and {Mumford}, S.~J. and {Muna}, D. and {Murphy}, N.~A. and {Nelson}, S. and {Nguyen}, G.~H. and {Ninan}, J.~P. and {N{\"o}the}, M. and {Ogaz}, S. and {Oh}, S. and {Parejko}, J.~K. and {Parley}, N. and {Pascual}, S. and {Patil}, R. and {Patil}, A.~A. and {Plunkett}, A.~L. and {Prochaska}, J.~X. and {Rastogi}, T. and {Reddy Janga}, V. and {Sabater}, J. and {Sakurikar}, P. and {Seifert}, M. and {Sherbert}, L.~E. and {Sherwood-Taylor}, H. and {Shih}, A.~Y. and {Sick}, J. and {Silbiger}, M.~T. and {Singanamalla}, S. and {Singer}, L.~P. and {Sladen}, P.~H. and {Sooley}, K.~A. and {Sornarajah}, S. and {Streicher}, O. and {Teuben}, P. and {Thomas}, S.~W. and {Tremblay}, G.~R. and {Turner}, J.~E.~H. and {Terr{\'o}n}, V. and {van Kerkwijk}, M.~H. and {de la Vega}, A. and {Watkins}, L.~L. and {Weaver}, B.~A. and {Whitmore}, J.~B. and {Woillez}, J. and {Zabalza}, V. and {Astropy Contributors}},
	doi = {10.3847/1538-3881/aabc4f},
	eid = {123},
	eprint = {1801.02634},
	journal = {\aj},
	month = sep,
	number = {3},
	pages = {123},
	primaryclass = {astro-ph.IM},
	title = {{The Astropy Project: Building an Open-science Project and Status of the v2.0 Core Package}},
	volume = {156},
	year = 2018}

@article{2013A&A...558A..33A,
	adsurl = {https://ui.adsabs.harvard.edu/abs/2013A&A...558A..33A},
	archiveprefix = {arXiv},
	author = {{Astropy Collaboration} and {Robitaille}, Thomas P. and {Tollerud}, Erik J. and {Greenfield}, Perry and {Droettboom}, Michael and {Bray}, Erik and {Aldcroft}, Tom and {Davis}, Matt and {Ginsburg}, Adam and {Price-Whelan}, Adrian M. and {Kerzendorf}, Wolfgang E. and {Conley}, Alexander and {Crighton}, Neil and {Barbary}, Kyle and {Muna}, Demitri and {Ferguson}, Henry and {Grollier}, Fr{\'e}d{\'e}ric and {Parikh}, Madhura M. and {Nair}, Prasanth H. and {Unther}, Hans M. and {Deil}, Christoph and {Woillez}, Julien and {Conseil}, Simon and {Kramer}, Roban and {Turner}, James E.~H. and {Singer}, Leo and {Fox}, Ryan and {Weaver}, Benjamin A. and {Zabalza}, Victor and {Edwards}, Zachary I. and {Azalee Bostroem}, K. and {Burke}, D.~J. and {Casey}, Andrew R. and {Crawford}, Steven M. and {Dencheva}, Nadia and {Ely}, Justin and {Jenness}, Tim and {Labrie}, Kathleen and {Lim}, Pey Lian and {Pierfederici}, Francesco and {Pontzen}, Andrew and {Ptak}, Andy and {Refsdal}, Brian and {Servillat}, Mathieu and {Streicher}, Ole},
	doi = {10.1051/0004-6361/201322068},
	eid = {A33},
	eprint = {1307.6212},
	journal = {\aap},
	month = oct,
	pages = {A33},
	primaryclass = {astro-ph.IM},
	title = {{Astropy: A community Python package for astronomy}},
	volume = {558},
	year = 2013}

@article{2024ApJ...974..299Y,
	adsurl = {https://ui.adsabs.harvard.edu/abs/2024ApJ...974..299Y},
	archiveprefix = {arXiv},
	author = {{Yoon}, Yongmin and {Kim}, Jae-Woo and {Ko}, Jongwan},
	doi = {10.3847/1538-4357/ad7816},
	eid = {299},
	eprint = {2410.17862},
	journal = {\apj},
	month = oct,
	number = {2},
	pages = {299},
	primaryclass = {astro-ph.GA},
	title = {{The Most Massive Early-type Galaxies Exhibit Tidal Features More Frequently in Lower-density Environments}},
	volume = {974},
	year = 2024}
\bibliographystyle{aasjournalv7}



\end{document}